\documentclass[prc,twocolumn,showpacs,preprintnumbers]{revtex4}
\usepackage{bm}
\usepackage{amssymb}
\usepackage{amsmath}
\usepackage{graphicx}
\begin{document}

\title{Black Hole Accretion Disk Diffuse Neutrino Background}
\author{T. S. H. Schilbach}
\author{O. L. Caballero}\email{ocaballe@uoguelph.ca}
\affiliation{Department of Physics,
             University of Guelph, Guelph, ON N1G 2W1, Canada }
\author{G. C. McLaughlin}%\email{gail\_mclaughlin@ncsu.edu} 
\affiliation{Department of Physics,
             North Carolina State University, Raleigh, NC 27695}

%%%%%%%%%%%
\date{\today}
\begin{abstract}
We study the cosmic MeV neutrino background from accretion disks formed during collapsars and the coalescence of compact-object mergers.
We provide updated estimates, including detection rates, of relic neutrinos from collapsars, as well as estimates for neutrinos that are produced in mergers.
Our results show that diffuse neutrinos detected at HyperK 
would likely include some that were emitted from binary neutron-star mergers.
The collapsar rate is uncertain, but at its upper limit relic neutrinos from these sources would provide a significant contribution to the Cosmic Diffuse Neutrino Background.
\end{abstract}
\smallskip
\maketitle

%%%%%%%%%%%%%%%%%%%%%%%%%%%%%%%%%%%%%%%%%%%%%%%%%%%%%%%%%%%%%%%%%%%%%%%%%%%%%%%
\section{Introduction}
%%%%%%%%%%%%%%%%%%%%%%%%%%%%%%%%%%%%%%%%%%%%%%%%%%%%%%%%%%%%%%%%%%%%%%%%%%%%%%%

%Black hole (BH) 
Accretion disks surrounding black holes (BH) or hypermassive neutron stars (HMNS) are likely the final fate of the coalescence of a neutron star (NS) with a compact object (BH or NS) \cite{Lee1999,Rosswog:2005su,Foucart:2014nda,Fryer:2015uia,Lehner:2016lxy}. Accretion disks are also
formed during rare supernovae that have significant rotation, termed  
collapsars \cite{MacFadyen:1998vz,OConnor:2010moj,Ott:2010gv,Sekiguchi:2010ja}. 
During these events, much of the gravitational energy is released as neutrinos. The neutrinos are interesting not only because of their key role in the setting of the
electron fraction and subsequent synthesis of elements, e.g. \cite{Surman_Mclaughlin04,Surmanrprocess,Surman:2008qf,Caballero:2011dw,Just:2014fka}, or their suspected
contribution to the triggering of long duration gamma ray bursts (see e.g. \cite{Woosley:1993wj,Popham1999,Kneller:2004jr,Murguia-Berthier:2016fys}), but also 
because they are one of the signals that come from these multi-messenger objects.  
Even from a small number of neutrinos (just like in the SN1987 case \cite{Hirata:1987hu}),
much can be gleaned about the central engines of these objects.

Neutrinos emitted from these accretion disks are expected to be in the energy range of MeV.  It is well known that astrophysical MeV neutrinos
could be registered at existing facilities such as SuperKamiokande (SK) \cite{SK} and SNOLAB \cite{Descamps:2015hva}.
The prospects of detection in larger facilities like the proposed Hyperkamiokande, UNO, DUNE, JUNO and TITAND \cite{hk,uno,Acciarri:2015uup,An:2015jdp,Titand}, are even more promising \cite{Kistler:2008us}. 
There are two basic strategies for detecting these MeV neutrinos, either a direct detection from an object that is sufficiently close to produce a substantial flux at earth, or a detection of the cosmic MeV neutrino background.
The latter is formed by the accumulation of neutrinos from such objects over time.
The consideration of the cosmic MeV neutrino background (CMNB) from all types of extra galactic
sources (supernovae, collapsars, binary mergers)
 enhances our chances of detection and 
opens a window to
neutrino physics at cosmological scales. A detection of the CMNB will provide
insights to the star formation history, initial mass function, cosmic metallicity, and
event rates (see e.g. \cite{Nakazato:2015rya,Davis:2017mbq}).
 
While many types of events can contribute to the CMNB, two types of events have been explored most extensively.   
Due to its promising prospects for detection, the supernova relic neutrino (SRN) background has been widely studied 
(e.g \cite{Kaplinghat:1999xi,Beacom:2003nk,Strigari,Ando2003,Ando2004,Lunardini2006,Lunardini:2010ab}, for reviews see  e.g. \cite{Beacom:2010kk,Lunardini:2010ab}).
SRN searches at SK have significantly improved 
upper limits, and they are now 
very close to predictions \cite{Malek:2002ns,Bays:2011si,Zhang:2013tua}.
The next most studied contribution to the CMNB is 
that of relic neutrinos
from failed supernovae (or unnovae).  Theoretical fluxes 
\cite{Lunardini2009failed,Keehn:2010pn,Yang:2011xd,Priya:2017bmm,Nakazato:2015rya}  have been found 
to be comparable to that of supernovae \cite{Nakazato:2008vj}. 

In this paper we add to previous CMNB studies by considering the neutrino background due to accretion disks from compact object mergers and supernovae (collapsars). We use updated models to extend previous work on 
collapsars, for example that of  \cite{Nagataki:2002bn} which found 
optimistic detection prospects for TITAND, using 
a neutrino background 
determined from the collapsar model in \cite{Nagatakicounts}. 
We also make the first  determination of the diffuse neutrino background from compact object mergers.
In both scenarios, matter surrounding
the remnant black hole or hypermassive neutron star is hot and will emit  considerable numbers of neutrinos.
The study of the accretion tori allows for a determination of neutrino emission, \textit{after} black hole formation, of collapsars and mergers.
By considering two different accretion disk models, discussed later, 
we investigate the effect that the accretion rate and the BH spin have on the
neutrino spectra, on the relic background, and on the associated number of neutrinos reaching Earth's detectors. 

The derivation of an accretion disk diffuse flux relies on two components: the neutrino spectra emitted in
one of the above scenarios
and the cosmological rate at which these events occur.
In both collapsars and mergers, the neutrino emission can be comparable to or even larger than 
supernovae.
In the collapsar case, simulations have shown
that the neutrino emission may be larger than that of a proto-neutron star \cite{Fischer:2008rh,Sumiyoshi:2008zw}. 
In the case where the disk is formed after a merger, the neutrino emission from one event, although shorter in
duration, can be one or two orders
of magnitude more luminous than that of a supernova \cite{Cab2009}.
 Similar to \cite{Nagataki:2002bn}  we employ 
steady state models of accretion disks
where the disk is considered to be the result of a collapsar, but in addition we consider a dynamical model. 
Our estimates come from updated models which include neutrino cooling, a range of black hole spin and estimates 
of gravitational bending and redshifting on the part of the neutrinos. We assume that the BH has been already formed and matter, in a torus distribution, is accreted into it at a given rate.  We also comment on the case of an accretion disk surrounding a hypermassive neutron star.

The other component, the cosmological failed supernova and merger rates, 
has been revisited in the recent years, motivating also this study.
From one side the detection of gravitational waves from 
mergers at observatories such as Advanced-LIGO, has triggered an impressive effort to estimate the merger coalescence rates \cite{OShaughnessy:2008yul,dominik}; with estimates in the range of  ($10^{-6}-10^{-3}$/year per Milky Way Equivalent Galaxy (MWEG) 
for NS-NS mergers, and $10^{-8}-3\times10^{-5}$/year per MWEG for BH-NS mergers \cite{Abadie:2010cf}).
The recent detection of a neutron star mergers, suggests a rate of $1540^{+3200}_{-1220}$  Gpc$^{-3}$yr$^-1$ \cite{TheLIGOScientific:2017qsa}. On
the other side, recent \textit{Swift} gamma rays burst data \cite{Gehrels:2004aa,Butler:2007hw} has been used to provide new estimates for star formation rates \cite{Yuksel2008}
and failed supernovae \cite{Yuksel:2012zy}.

In this manuscript we convolve the accretion disk neutrino spectra from two different models, with current failed supernova and merger rates.
In doing so, we  provide an updated baseline for future studies on relic neutrinos from collapsars, the first estimates from mergers, and comparison between the two scenarios.  We focus on the
electron antineutrino relic flux, its contribution to the MeV neutrino background, and its possible
detection at water Cherenkov facilities.

Although important, we do not consider neutrino oscillations in this work. Oscillations
will change the large energy contribution of the neutrino spectra resulting in a larger
number flux of relic electron antineutrinos (see e.g. \cite{Ando2003,Lunardini:2012ne}). 
Oscillations are expected to play a significant role in mergers and collapsars, e.g. \cite{Malkus:2012ts,Malkus:2014iqa,Zhu:2016mwa} and we will discuss 
the role of oscillations in the accretion disk relic neutrinos in future work.

This paper is organized as follows: in section \ref{disk model} we discuss the
accretion torus models used and in section \ref{spectra} we present the
corresponding results for the neutrino
spectra. We continue by introducing
the compact object mergers and failed supernovae rates used in this work and show our results
for the relic neutrino flux for each scenario in section \ref{diffuse flux}. 
In section \ref{detection rates} we provide neutrino event rates at SK and finally
in section \ref{conclusions} we conclude.
%%%%%%%%%%%%%%%%%%%%%%%%%%%%%%%%%%%%%%%%%%%%%%%%%%%%%%%%%%%%%%%%%%%%%%%%%%%%%%

%%%%%%%%%%%%%%%%%%%%%%%%%%%%%%%%%%%%%%%%%%%%%%%%%%%%%%%%%%%%%%%%%%%%%%%%%%%%%%%
\section{Disk Models}
\label{disk model}
%%%%%%%%%%%%%%%%%%%%%%%%%%%%%%%%%%%%%%%%%%%%%%%%%%%%%%%%%%%%%%%%%%%%%%%%%%%%%%%
The diverse emissions that could be observed from binary mergers and collapsars,
have significantly stimulated the study of accretion disks in the last decades. Models 
incorporate neutrino cooling and 
utilize a variety of different methods which include treatments that are
fully relativistic, Newtonian, hydrodynamical, steady state and dynamical; a few examples include
e.g. \cite{Setiawan04,Sekiguchi:2010ja,Foucart:2014nda,Perego:2014fma}. In this work
we make use of two different disks models. One is a fully relativistic steady-state model by Chen and Beloborodov \cite{Beloborodovcross},
and the other one is from a pseudo-relativistic hydrodynamical simulation from Just et al. \cite{Just:2014fka}. 
Based on these two models we calculate the neutrino spectra and the corresponding diffuse background, aiming to set  
bounds on the number of neutrino events detected on Earth.
We briefly summarize these two models below.

In the first model, from Chen and Beloborodov, the disk is arranged to depend on radius solely.
We extend it to 
3D by assuming axial symmetry and estimating 
the vertical structure with a simple hydrostatic model. The disk is formed by a gas 
of nucleons, $\alpha$-particles, electrons, positrons, photons, and neutrinos in nuclear statistical equilibrium (NSE). Both the gas and radiation pressures are at equilibrium. 
The model is fully relativistic and uses the Kerr metric to account for two values of the BH spin $a=Jc/GM^2=0$ and $0.95$ ($J$ is the total angular momentum and $M$ the BH mass). 
The effects of three-dimensional magneto-hydrodynamical turbulence are approximated as is usual via one viscosity parameter $\alpha$ \cite{Shakura:1972te}. 
In what follows, these models are labelled according to the BH spin: ``C0'' for $a=0$ and ``Ca'' for $a=0.95$. 
The mass of the BH is 3$M_\odot$ and the alpha viscosity given by $\alpha = 0.1$.
Steady accretion is assumed, allowing us to study the effect of a constant mass accretion rate, $\dot M$, on the neutrino spectra.
For this model, we have used values of $\dot{M}$= 3, 5, 7 and 9 $M_\odot$/s. 
Observations of short and long gamma ray bursts luminosities suggest a range of accretion  values between
0.1 -10 $M_\odot$ \cite{Shapiro:2017cny}.
Fully relativistic dynamical simulations of high entropy rotating stellar cores 
show that the accretion rate just before  and after BH formation varies depending on the degree of rotation and may be as high as
45 $M_\odot$/s, although decrease with time to values of 5 $M_\odot$/s \cite{Sekiguchi:2010ja}. Binary NS mergers simulations
find that the accretion rate can be 0.1-1 $M_\odot$/s \cite{Shibata:2006nm}, or as large as 10 $M_\odot$/s \cite{Shibata:2005mz}. 
Dynamical magnetized BHNS
mergers simulations have found  that the rate vary between 0.1 to around 5 $M_\odot$ /s \cite{Nouri:2017fvh}, while  \cite{Fernandez:2013tya, Deaton:2013sla}
found accretion rates around 1 $M_\odot$/s.

For the second model, we use the simulation results of Just et al. \cite{Just:2014fka} who studied the disk evolution, 
based on parameters extracted from hydrodynamical simulations of NS-NS and BH-NS mergers. Their work assumes the only merger result is BH-torus systems. 
The simulations are performed in Newtonian hydrodynamics and assume axisymmetry while ignoring the torus self-gravity due to its insignificance relative to the BH.
Relativistic effects are introduced by using the Artemova-potential \cite{Artemova} to describe the BH gravitational field, with the BH spin and mass held fixed. 
The equation of state assumed the same particles as the Chen and Beloborodov model above, but included a heavy nucleus ($^{44}$Mn), all of them in NSE. 
The simulations in \cite{Just:2014fka} begin with a BH and torus accreting onto it. 
The BH mass is  $3 M_{\odot}$ and the alpha viscosity was taken to be $\alpha = 0.02$. While the model describes the time evolution of the disk, however, we focus here on a representative time of $t=20$ ms.  
In the framework of this torus model, two BH spins are considered $a=0$, and $0.8$. These models are labelled in this work according to the BH spin: ``J0'' for $a=0$ and ``Ja'' for $a=0.8$.

While neutrino cooling is already included in the two models, we aim to calculate neutrino spectra and diffuse fluxes for distant observers, therefore, 
we use our results from previous work, where we performed a ``post-processing'' of the tori's thermodynamical properties and found
the last points of neutrino scattering a.k.a the neutrino sphere. Details on the calculation and discussion on the results can be found in \cite{Cab2009,Caballero:2015cpa}.

In the case of binary neutron-star mergers, 
fully general-relativistic simulations of have shown that rapidly rotating merger remnants allow
the formation of a HMNS (see e.g \cite{Hotokezaka:2013iia}). The lifetime of this HMNS depends at least on angular momentum transport, gravitational
wave emission, the equation of state, and neutrino cooling.
When the angular momentum transport is dominant the HMNS will collapse into a BH otherwise it will collapse after neutrino cooling. 
The neutrino luminosities of the HMNS found by Sekiguchi et al \cite{Sekiguchi:2011zd} with relativistic simulations,
and by Lippuner et al \cite{Lippuner:2017bfm}, who studied HMNS with their accretion disks and their
evolution after collapse to BHs in pseudo-Newtonian gravity, are of the order of $10^{53}$ ergs/s. 
The simulations shown that neutrino emission will continue after collapse decreasing from the initial values
set by the HMNS. The order of magnitude of the luminosities is the same as
our results for BH-AD used here \cite{Caballero:2015cpa}. 

%%%%%%%%%%%%%%%%%%%%%%%%%%%%%%%%%%%%%%%%%%%%%%%%%%%%%%%%%%%%%%%%%%%%%%%%%%%%%%%
\section{Neutrino spectra}
\label{spectra}
%%%%%%%%%%%%%%%%%%%%%%%%%%%%%%%%%%%%%%%%%%%%%%%%%%%%%%%%%%%%%%%%%%%%%%%%%%%%%%%

Neutrinos produced in accretion disks (and supernovae)  are trapped due to the highly dense matter
of these environments. The neutrinos begin free streaming regime at the neutrino sphere, and therefore 
the neutrino properties observed at any point in space above the neutrino surface are characterized by the thermodynamical properties of such surface.
To calculate the neutrino spectra we consider the number of neutrinos that are emitted from a mass element on the neutrino surface of a black 
hole accretion disk (BH-AD) with energy $E$. 
The number spectrum of neutrinos emitted by one BH-AD is

\begin{equation}
\frac{dN(E)}{dE} = \frac{g_\nu c}{2\pi^2(\hbar c)^3}\iint dA f(E) dt,
\label{numspec}
\end{equation}
where $dt$ corresponds to the total emission time and $g_\nu = 1$. 
We assume that the emission of neutrinos by one mass element is isotropic and with the integral over the area we sum over all mass elements. This integral is expanded as

\begin{equation}
\iint dA = \int^{2\pi}_0d\phi\int^{\rho_{2}}_{\rho_{1}}\rho d\rho,
\end{equation}
where $\rho_{2}$, $\rho_{1}$ and $\phi$ are respectively the outer radius, inner radius, and angular component of the neutrino surface in cylindrical coordinates. The function $f(E)$ in Eq. \ref{numspec} is the usual Fermi-Dirac distribution for fermions:

\begin{equation}
f(E) = \frac{E^2}{e^{E/T}+1}, 
\end{equation}
with $T$ the \textit{local} temperature at the neutrino surface \cite{Ando2004}.

As described up to this point, Eq. \ref{numspec} corresponds to the neutrino spectrum as seen by a local observer ($E$ and $T$ as seen in the comoving frame of the disk). 
However, the spectrum observed at a distant point, $dN(E_\infty)/dE_\infty$, is affected by relativistic effects due to the strong gravitational field of the BH such as 
energy shifts, time dilation, and bending of neutrino trajectories. This means that if a disk is emitting neutrinos in a galaxy and is observed several kpc
away the energy and time are redshifted as $E = (1+z_{BH})E_\infty$, $dt=(1+z_{BH})dt_\infty$ and the toroidal neutrino 
surface would appear larger by  $dA_\infty=(1+z_{BH})^2 dA$, with $(1+z_{BH})$ the redshift due to the BH. 
For the two models used here, steady-state and hydrodynamical, we use estimates found in previous work (noted there as $\Delta t_\infty$) of the total emission time $dt_\infty$ \cite{Caballero:2015cpa}. 
Those time intervals are based on the efficiency to convert gravitational binding energy into neutrino energy by 
considering its dependency with the BH spin. 

\begin{figure}[ht]
\begin{center}
\includegraphics[width=0.45\textwidth] {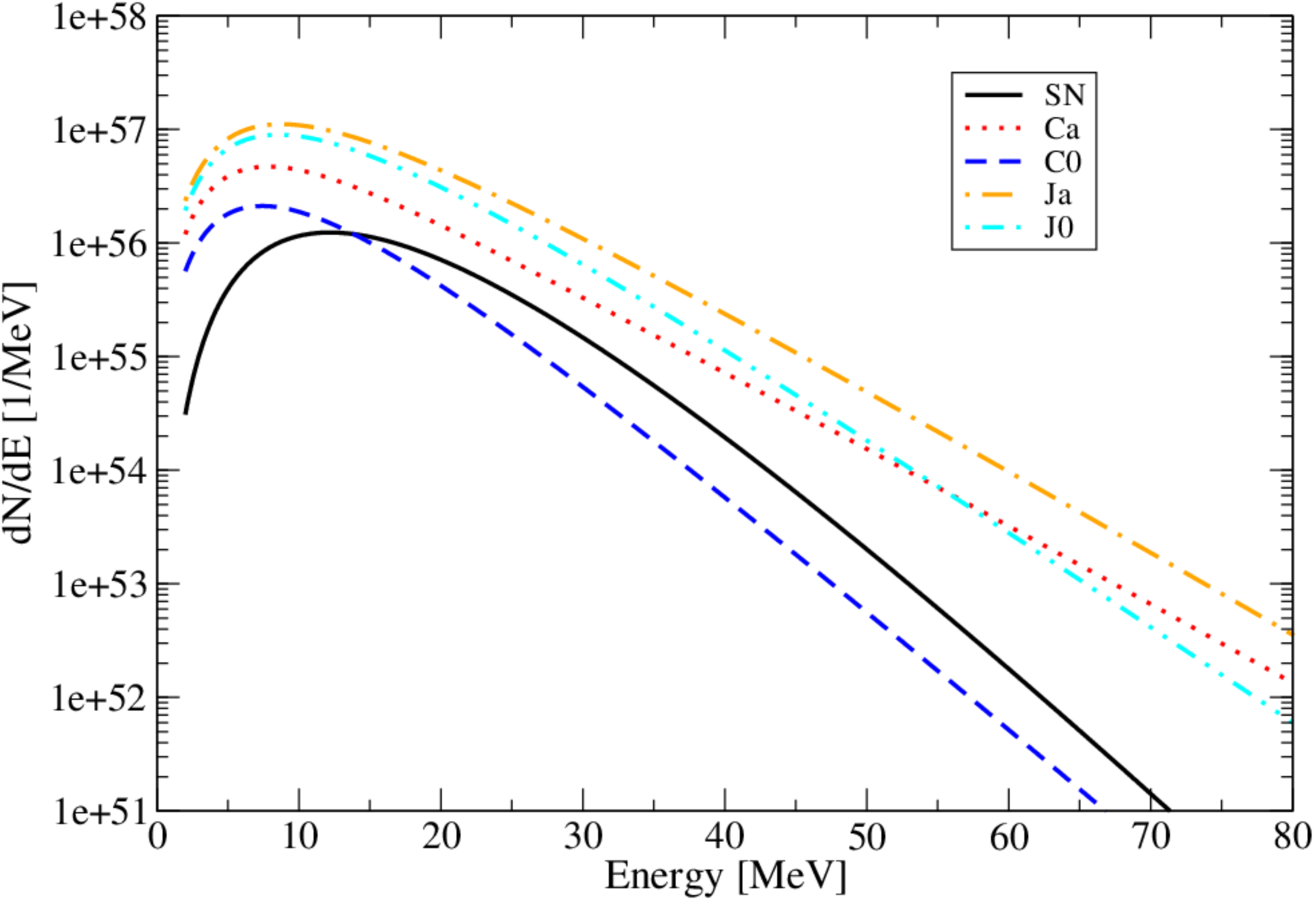}
\caption{Comparison of electron antineutrino spectra for the steady-state (C0, Ca  with constant accretion rate $3M_\odot$/s) and dynamic tori (J0, Ja).
The $a$ indicates a large BH spin (see text). In both disk models the BH mass is $3 M_\odot$. The supernova spectrum (SN) is shown for comparison.}

\label{spectrum}
\end{center}
\end{figure}

Taking into account the above strong gravitational field effects and the conservation of phase-space density, we generate the spectra observed far away from the source.
In what follows we focus on electron antineutrinos only as we aim to calculate detection rates at water-based Cherenkov detectors.
Fig. \ref{spectrum} shows the results for electron antineutrinos using the steady-state and the hydrodynamical tori and compares with a protoneutron star spectrum \cite{Ando2003,Totani:1997vj}. 
In Fig. \ref{spectrum} we have used an accretion rate of $\dot M= 3 M_\odot$/s for the C$a$ and C0 models. As can be seen, the 
disk spectra are larger than the supernova (SN) one (from \cite{Totani:1997vj}), except for the C0 model which is larger only for
energies below $E<14 $ MeV.
These results are consistent with the fact that accretion disks formed during mergers may result in higher neutrino emission temperatures than those 
of SN. 
On the other hand, if the disk is a result of a collapsar
the results are also consistent with the fact that the formation of a BH (instead of a NS) 
creates a situation where, 
depending on the accretion rate, significant energy can become available for conversion to neutrino
emission. 
In the C0 model with $\dot M= 3 M_\odot$/s, although the highest neutrino temperature is $T=4.3$ MeV (close to the well known SN value of 5 MeV), 
the range of temperatures at the neutrino
surface is below  that of a SN. In contrast, at the same
accretion rate, the change to a spining BH (C$a$ model, described in the Kerr metric) will generate hotter disks with larger
angular momentum and inner edges closer to the BH where most of the power is released. The energy is transferred 
via viscous heating and then converted to neutrino energy.
Similarly, for the hydrodynamical models, we find that the J$a$ torus has a larger neutrino spectrum compared to the J0 model, particularly at
high energies.

\begin{figure}[ht]
\begin{center}
\includegraphics[width=0.45\textwidth] {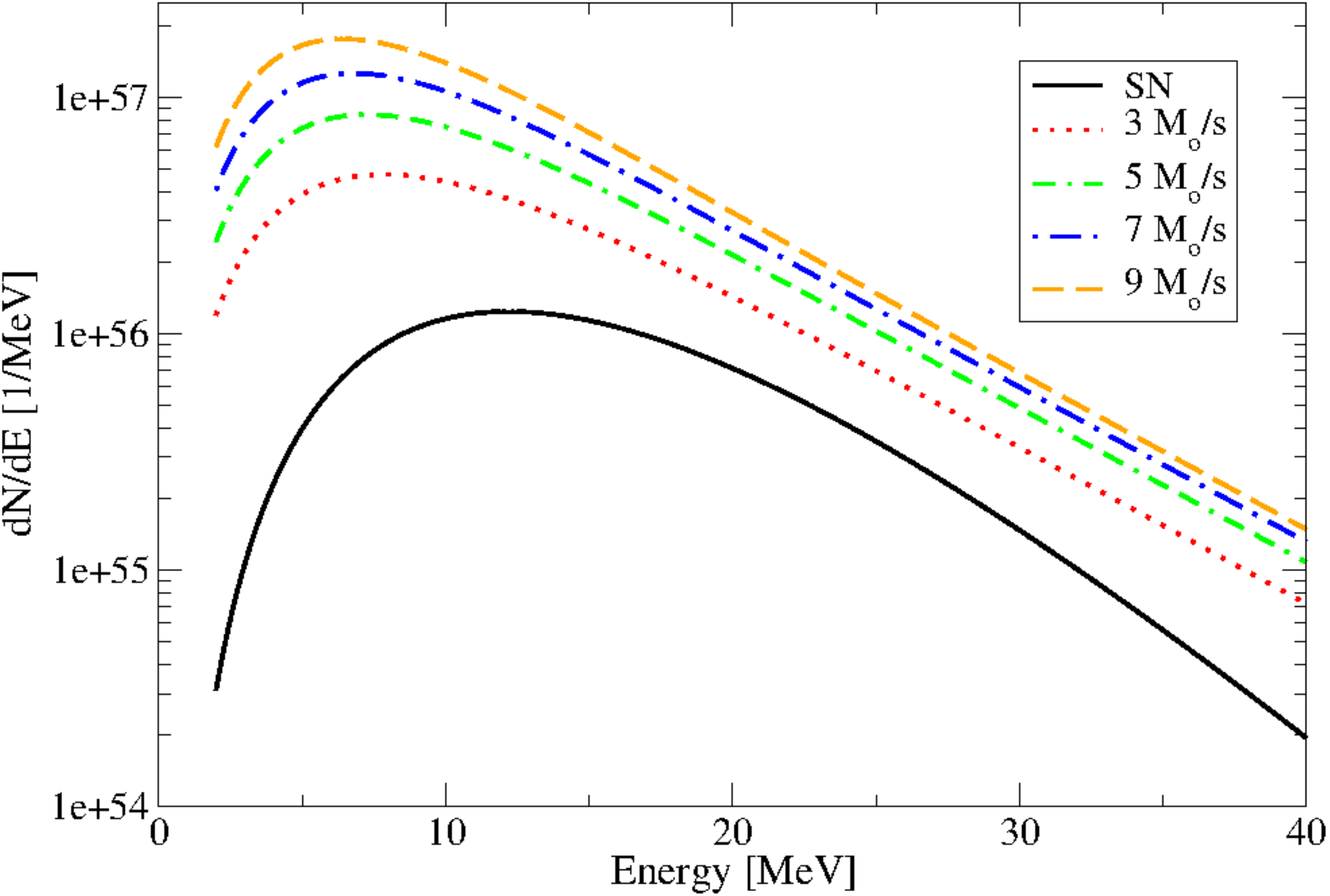}
\caption{The effect of mass accretion rate on the AD electron antineutrino spectra for the steady-state Ca model (BH spin $a=0.95$).}
\label{spectrumRateSpread}
\end{center}
\end{figure}

In the case of accretion rate dependence, as shown in Figure \ref{spectrumRateSpread} for the C$a$ models,
a similar conclusion is drawn: the larger the rate at which  mass plunges into the BH 
the higher the temperature of disk \cite{Beloborodovcross}
and, therefore, the number of 
antineutrinos emitted per energy interval.
As a result, a torus 
with a high accretion rate will have a 
stronger signal
than a slower accreting disk \cite{Caballero:2015cpa}.
 
Finally, comparing our results to those of Ref. \cite{Nagataki:2002bn,Nagatakicounts} who use a collapsar model, we find that their neutrino emission is
smaller.
It is important to note that in their work, the authors modeled the
disk as an advection-dominated flow, whereas the models presented here both allow for a neutrino-dominated phase. Also, other parameters such as
a smaller accretion rate ($0.1 \dot{M}_\odot$/s), and lower temperatures contribute to the differences. This, of course, will have implications 
on the diffuse neutrino flux as discussed later. 

The higher temperatures of accretion disk tori are similar to those from
failed SN \cite{Nakazato:2013maa}. However, the overall magnitude of the failed SN
spectra of \cite{Nakazato:2013maa}  is smaller than our collapsar results.
This is primarily because that case involves 
spherically symmetric matter distributions and neutrino emission only up to the onset of the BH, whereas in the models we employ 
the disk evolution and its neutrino emission correspond to the period after BH formation.

%%%%%%%%%%%%%%%%%%%%%%%%%%%%%%%%%%%%%%%%%%%%%%%%%%%%%%%%%%%%%%%%%%%%%%%%%%%%%%%
\section{Diffuse flux}
\label{diffuse flux}
%%%%%%%%%%%%%%%%%%%%%%%%%%%%%%%%%%%%%%%%%%%%%%%%%%%%%%%%%%%%%%%%%%%%%%%%%%%%%%%

One single torus emits neutrinos according to the spectra found in the previous section.
To find the disk diffuse neutrino background, we should consider the total number of disks that have emitted neutrinos from the past to the present time,
and convolve it with the cosmologically redshifted neutrino spectrum.
The number of disks, formed at a fixed time in the past, depends
on the event rate $R(z_{C})$ (number density of scenarios ending in a torus per unit time), which changes
with the cosmological redshift $z_C$. This rate has to be transformed to account for the
expansion of the Universe. In this way, we have
that the present number density of BH-AD neutrinos, observed now between the energy interval $E_o+dE_o$, emitted in the redshift interval $z_C+dz_C$ is given by 

\begin{equation}
\frac{dn_\nu(E_o)}{dE_o}=(1+z_C)R(z_C)\frac{dt_C}{dz_C}dz_C\frac{dN(E_\infty)}{dE_\infty},
\end{equation}
where $dN(E_\infty=(1+z_C)E_o)/dE_\infty$ is the number spectrum of neutrinos emitted by a single BH-AD, $E_o$ is the registered energy on earth and redshifted from $E_\infty$. The last two energies are related by $E_\infty=(1+z_C)E_o$. 

The Friedmann equation gives the relation between the past time $t_C$ and $z_C$ as

\begin{equation}
\label{dzdt}
\frac{dt_C}{dz_C}= -1 / (H_0(1+z_C)\left(\Omega_m(1+z_C)^3+\Omega_\Lambda\right)^{1/2}),
\end{equation}
where $\Omega_m=0.3$, $\Omega_{\Lambda}=0.7$ and $H_0$=70 km/s/Mpc in the $\Lambda$CDM standard cosmology (see e.g. \cite{Hogg:1999ad}).

The differential number flux of BH-AD diffuse neutrinos, $dF/dE_o=c dn/dE_o$, is obtained by summing over the cosmological redshift:

\begin{equation}
\label{flux}
\frac{dF}{dE_o}=-c \int^{z_{max}}_0 (1+z_C) R(z_C)\frac{dN(E_\infty)}{dE_\infty}\frac{dt}{dz_C} dz_C,
\end{equation}
with $dN(E_\infty)/dE_\infty$ the transformed spectrum in terms of the observed energy on Earth, $E_o$. Therefore, two primary red shifts were factored into this work, 
both from cosmological expansion and from the escape from the parent BH. 

A key ingredient in our calculation is the event rate $R(z_C)$, which changes with the cosmological redshift and depends on the scenario considered. 
Based on the BH-AD progenitor rate, $R(z_C)$, we calculate diffuse neutrino fluxes $dF/dE$ for the mergers and collapsar scenarios. 
Before presenting our results for the diffuse flux we discuss the event rates $R(z_C)$ used in this work.

%%%%%%%%%%%%%%%%%%%%%%%%%%%%%%%%%%%%%%%%%%%%%%%%%%%%%%%%%%%%%%%%%%%%%%%%%%%%%%%
\subsection{Compact object merger rates}
%%%%%%%%%%%%%%%%%%%%%%%%%%%%%%%%%%%%%%%%%%%%%%%%%%%%%%%%%%%%%%%%%%%%%%%%%%%%%%%

For our study of disk formation in the merger scenario
we use the results of Dominik \textit{et al} for extra-galactic compact object merger rates
(publicly available at http://syntheticuniverse.org) \cite{dominik}. The corresponding rates at $z_C=0$ 
are consistent with the lower limit inferred from the recent observation of gravitational waves from a NS-NS merger\cite{TheLIGOScientific:2017qsa}.
Should be noted that Dominik et al work corresponds to field stellar populations only and therefore their results, and the ones obtained here based on those, are a conservative
lower limit, as mergers occurring in globular clusters increase such rates. Their results for black hole-neutron star (BH-NS), neutron star-neutron star (NS-NS) 
are shown in Figure $3$ of \cite{dominik}, and we plot them here to provide context.
Their merger rates were broken into $4$ distinctive approaches to merger modeling: their standard baseline model, 
their optimistic common envelope model which allows for envelope donors, delayed SN model without a rapid SN engine, and high BH kicks with BHs providing full natal kicks.      

\begin{figure}[ht]
\begin{center}
\includegraphics[width=0.45\textwidth] {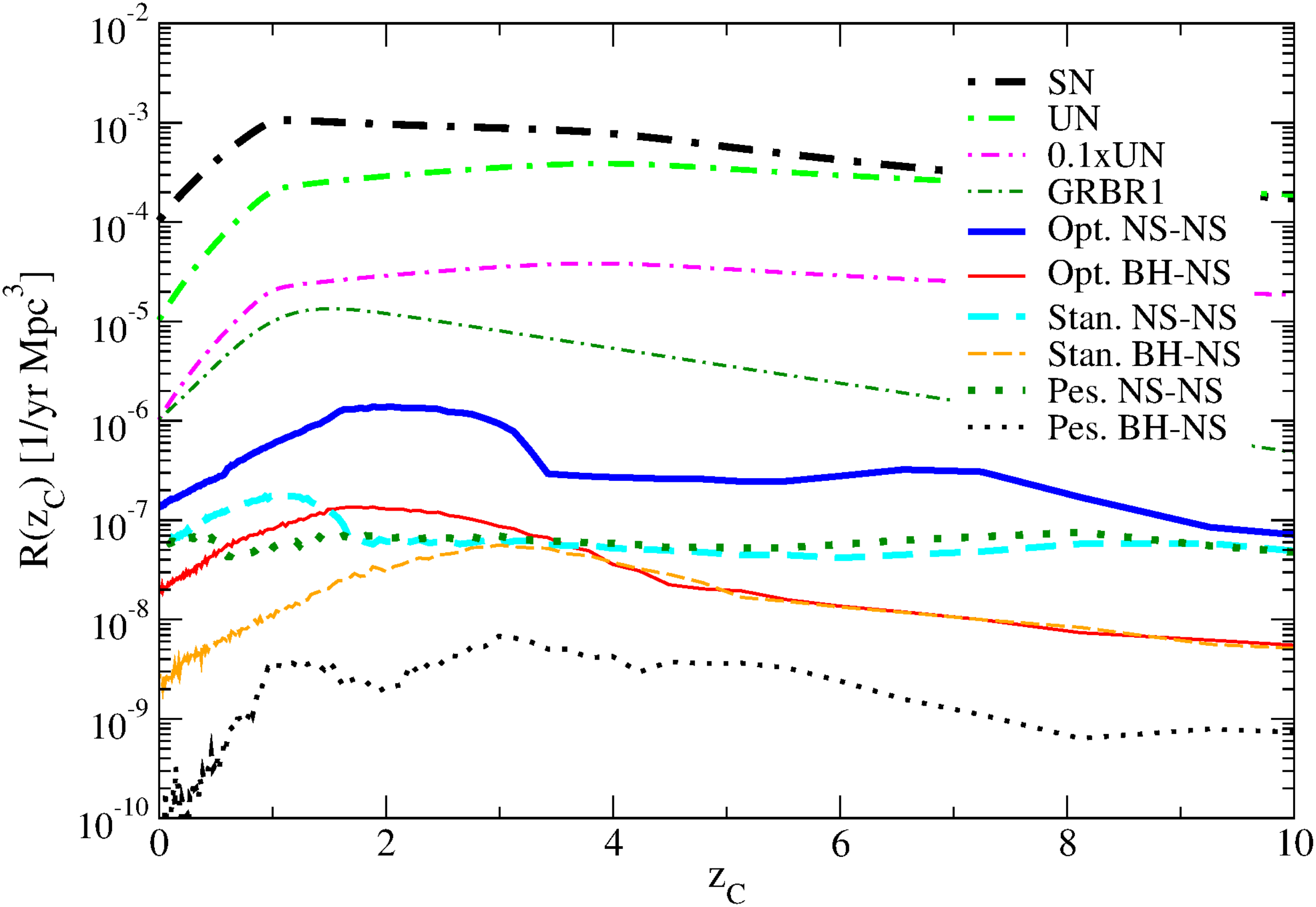}
\caption{Comparison of optimistic, standard and pessimistic cosmological BH-NS and NS-NS merger rates (from \cite{dominik}).
The SN and failed SN or unnova rate (UN) are as proposed in \cite{Yuksel:2012zy}. The 0.1xUN line assumes that only 10\%
of the UN would form a disk. GRBR1 is the collapsar rate as proposed in \cite{Nagataki:2002bn}.}
\label{ratedensity}
\end{center}
\end{figure}

For the purpose of analyzing the most optimistic and pessimistic cases within this data set, 
we plot upper and lower limits in Fig.  \ref{ratedensity}.
The upper limits for BH-NS and NS-NS correspond to the galactic low-end metallicity with common envelope merger scenario (labeled here
as Opt. BH-NS) and NS-NS  (Opt. NS-NS). The pessimistic cases are the high-end metallicity evolution scenario with BH kicks for BH-NS (in this work labeled as Pes. BH-NS)
and the low-end metallicity evolution scenario with the standard merger model of \cite{dominik} for NS-NS (here Pes. NS-NS). 
The two lines labeled as Stand. correspond to the high-end metallicity evolution in the standard model for BH-NS and NS-NS.

% 

%%%%%%%%%%%%%%%%%%%%%%%%%%%%%%%%%%%%%%%%%%%%%%%%%%%%%%%%%%%%%%%%%%%%%%%%
\subsection{Supernovae rates}
%%%%%%%%%%%%%%%%%%%%%%%%%%%%%%%%%%%%%%%%%%%%%%%%%%%%%%%%%%%%%%%%%%%%%%%

The SN rates are gathered from the results of Y\"{u}ksel and Kistler \cite{Yuksel:2012zy}, which
utilize an updated star formation history fit from Kistler et al. \cite{Kistler2013} based upon the original star formation rate fit done by Y\"{u}ksel et al. \cite{Yuksel2008}.

\begin{equation}
R = \zeta\dot{\rho_o}\left( (1+z_C)^{a\eta} + \left(\frac{1+z_C}{B}\right)^{b\eta} + \left(\frac{1+z_C}{C}\right)^{c\eta}\right)^{1/\eta},
\label{snrate}
\end{equation}
where $a=3.4$, $b=-0.3$, $c=-2$, $B= 5100$, $C= 14$, $\rho_o =0.014 M_{\odot}$/yr Mpc$^3$, and $\zeta = 0.0074/M_{\odot}$ \cite{Kistler2013,Yuksel:2012zy}.
To modify this rate for the particular case of failed SN, it is multiplied by
\begin{equation}
(1+z_C)/10,
\end{equation}
as 
discussed 
in Yuksel et al. \cite{Yuksel:2012zy}. This factor is a consequence of indications, given by bright gamma-ray bursts observations,
that the 
failed supernova rate may evolve
with a higher dependency of the cosmological redshift by a factor of $(1+z_C)$ \cite{Kistler:2013jza}.  
The factor scales with $z_C$ due to the theoretical expectation that lower metallicity stars will generate more massive cores 
The above parameterization of SN rates comes with the assumption that every star over $8 M_{\odot}$ experiences
a core collapse, and uses a Salpeter mass function that continues up until $100 M_{\odot}$, and further that 10\% of supernovae are unnovae and do not produce a supernova light curve.
We view this failed supernova rate as an upper limit to the number of supernova that could form accretion disks, as the 
fraction is relatively unconstrained. However, we must 
keep in mind that the true fraction could be smaller.
The 
failed SN rate is shown in Figure \ref{ratedensity} with the thick light-green dot-dashed line. 
Recent simulations have shown that massive stars with low metallicity and low mass loss can evolve in to a black hole with a disk \cite{Woosley:2011aw}.
Also, Sekiguchi et al \cite{Sekiguchi:2010ja} studied the evolution of high-entropy rotating stellar cores and found that even in the case
of a slowly rotating core the system evolves into a BH with a thin disk. 
Therefore, the fraction of black hole forming collapses that lead to a disk could be a substantial fraction of the failed
supernova rate. Nevertheless, to account for this uncertainty, we also provide estimates assuming that only 10 $\%$ of the failed SN would form a disk (magenta dot-dashed line).
For comparison the lowest estimate for collapsar rates of \cite{Nagataki:2002bn}, GRBR1, is also shown (thin dark-green dot-dashed line).

\subsection{Diffuse flux results}

\begin{figure}[ht]
\begin{center}
\includegraphics[width=0.45\textwidth] {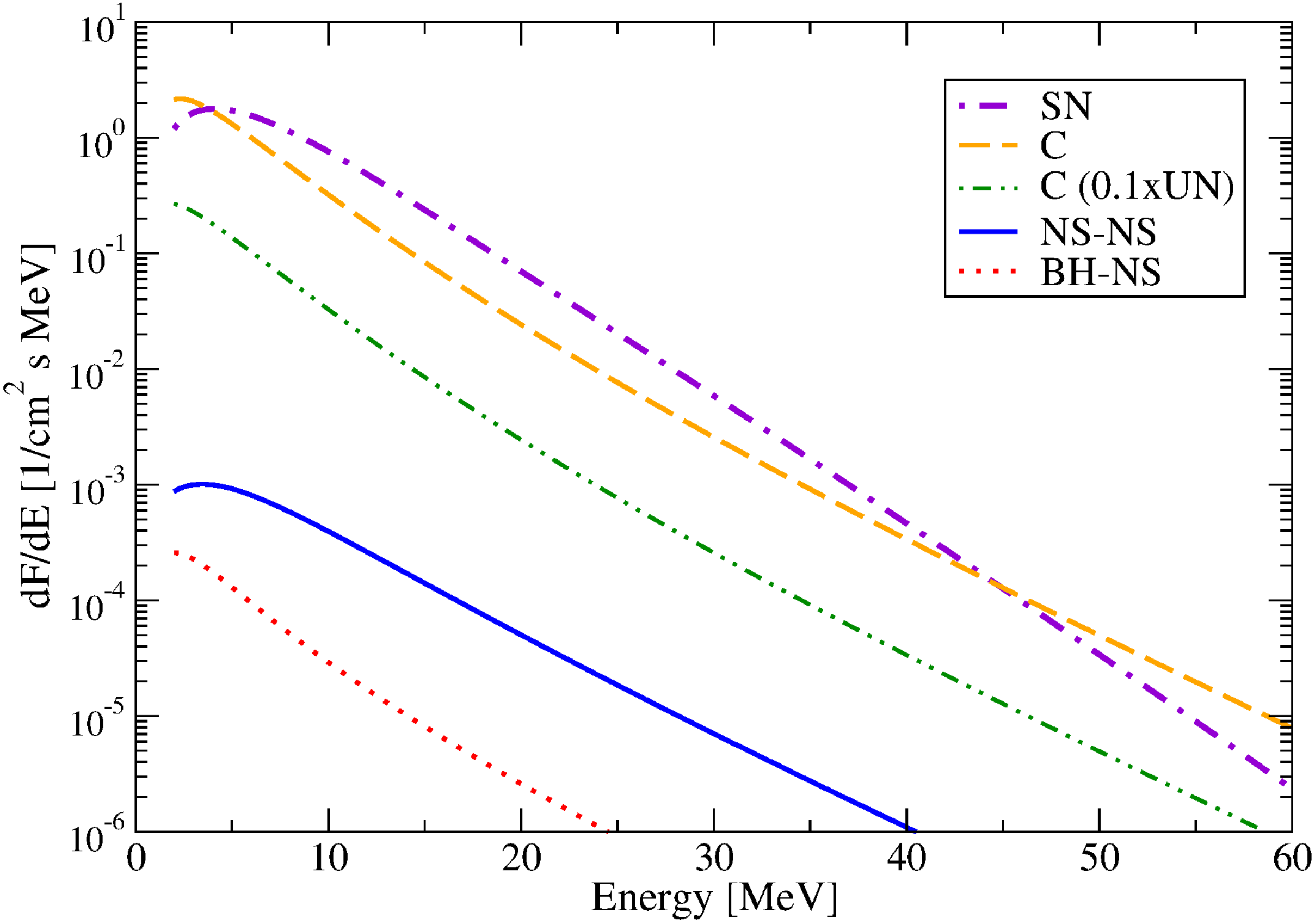}
\caption{Comparison of supernova (SN), collapsar upper limit (C), collapsar with 10$\%$ unnova rate (C(0.1$\times$UN)), BH-NS and NS-NS accretion disks diffuse neutrino fluxes 
where the accretion rate is $3 M_{\odot}$ / s and the BH spin is $a=0.95$ (steady-state model).}
\label{bhspec}
\end{center}
\end{figure}

We make estimates for the number flux of neutrinos based on eq. \ref{flux} for each astrophysical scenario. 
Figure \ref{bhspec} shows the diffuse fluxes for electron antineutrinos when the spectrum
corresponds to a disk with a $3 M_{\odot} / s$ accretion rate (Ca model). It can be seen that the upper limit for the  collapsar relic flux (dashed orange line)
is comparable to that of a SN (dotted-dashed black line). Based on our results of neutrino spectra (figures \ref{spectrumRateSpread} and \ref{spectrum}),
it follows that
the upper limit on the 
the collapsar diffuse background will be larger than the SN one for the $9 M_{\odot} / s$ Ca disk, and lower in the $3 M_{\odot} / s$ C0 case.
Therefore, for the accretion rates and BH spin ranges considered in this work, 
there exists the possibility that the number of neutrinos detected 
may be comparable to that of the diffuse SN background.
This is because, although the unnova rate is an order of magnitude lower than the SN rate, the binding energy 
available for neutrino emission in disks from collapsars is larger than the one in a SN.

Our results in the collapsar model for the diffuse background are significantly larger than those found by Nagataki et al
\cite{Nagataki:2002bn}. This in part due to an increased neutrino emission, and in part due to the using the unnovae rate as an upper limit on the collapsar flux (see Fig. \ref{ratedensity}).  
As discussed above the fraction of BH forming collapses
that evolve into a disk is still unknown.  However, if this fraction is not of orders of magnitude smaller than the unnovae,
then the diffuse neutrino flux in the collapsar scenario 
would contribute in a meaningful way to the CMNB. The green double dotted-dashed line in Figure \ref{bhspec} presents results where we have assumed
that only 10$\%$ of the failed supernovae would form a disk.

We also plot in Fig. \ref{bhspec} the contribution from 
NS-NS mergers and BH-NS mergers. As expected because of larger anticipated rates (see Fig.  \ref{ratedensity}) 
NS-NS mergers provide consistently greater contributions to the differential flux than the BH-NS, because of their 
larger rate of occurrence.

%%%%%%%%%%%%%%%%%%%%%%%%%%%%%%%%%%%%%%%%%%%%%%%%%%%%%%%%%%%%%%%%%%%%%%%%%%%%%%
\section{Detection rates}
\label{detection rates}
%%%%%%%%%%%%%%%%%%%%%%%%%%%%%%%%%%%%%%%%%%%%%%%%%%%%%%%%%%%%%%%%%%%%%%%%%%%%%%

The number of diffuse electron antineutrinos registered in a given facility per year, $R_D$, is obtained by integrating Eq. \ref{flux} with the detector 
cross section, $\sigma(E_o)$,

\begin{equation}
R_D=N_T\int_{E_{th}}{\sigma (E_o)\frac{dF}{dE_o}dE_o}.
\label{detection rate}
\end{equation}

Here $N_T$ is the number of targets in the detector, $E_{th}$ is its corresponding energy threshold, $E_o$ is the energy at the lab,
and $dF/E_o$ is the diffuse flux discussed in section \ref{diffuse flux}. 

For water-based Cherenkov detectors the relevant reaction is
\begin{equation}
\label{anuabsorption}
\bar{\nu}_e +p \rightarrow e^+ + n,
\end{equation}
where the cross section is given by
\begin{eqnarray}
\label{absorption}
\sigma_{{\bar\nu}_ep\rightarrow ne^+} &=& \frac{\sigma_0}{4m^2_e}(1+3g^2_A)(E_o-\Delta)^2\nonumber\\
&\times&\left[1-\left(\frac{m_e}{E_o-\Delta}\right)^2\right]^{1/2},
\end{eqnarray}
with $\sigma_0=4G^2_Fm^2_e/(\pi \hbar^4)$, $g_A = 1.26$, $m_e$ the electron mass, $\Delta = 1.293$ MeV the neutron-proton mass difference, and $G_F$ the Fermi coupling constant.

\begin{figure}[ht]
\begin{center}
\includegraphics[width=3.25 in,clip=true]
 {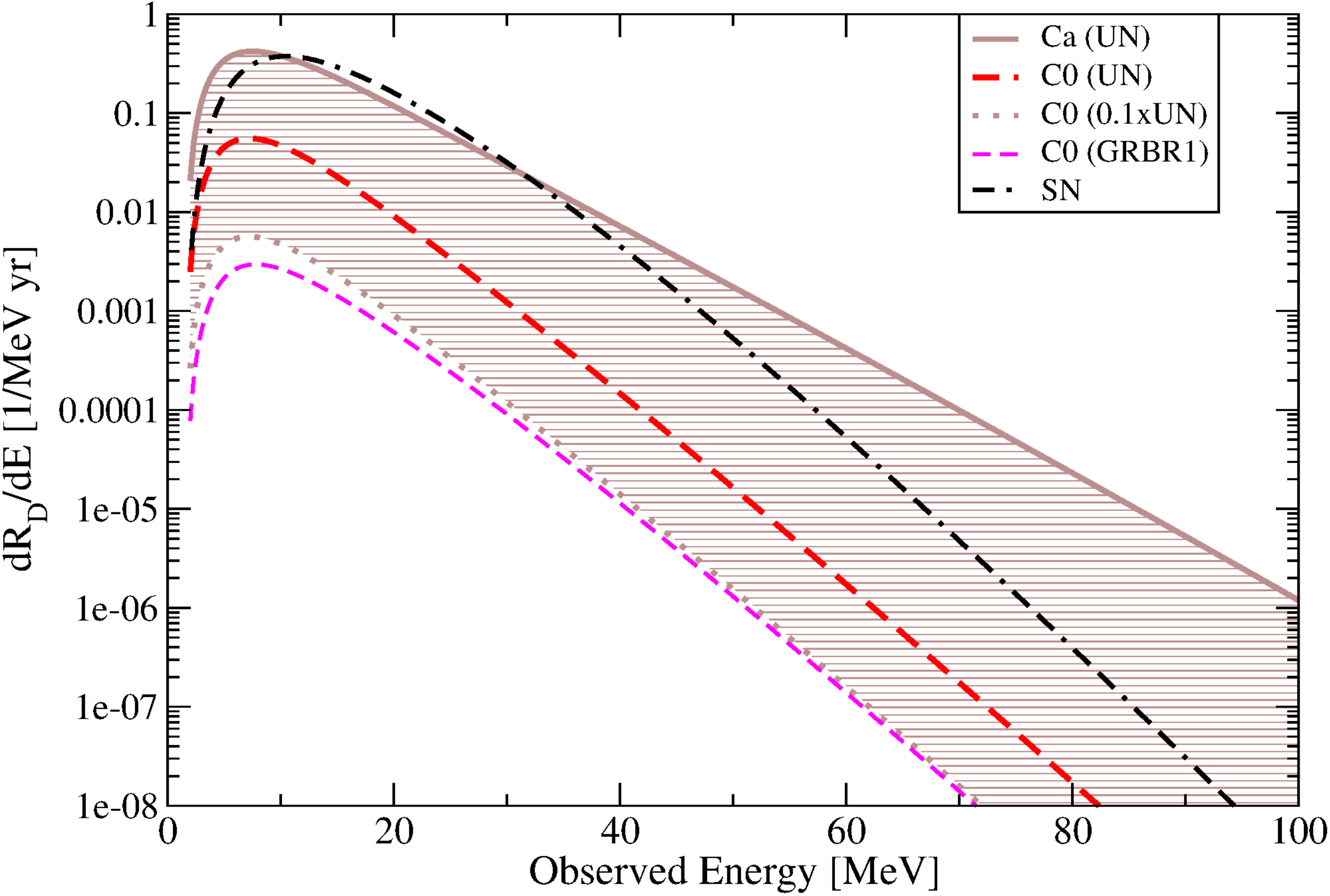}
\caption{(Color on line) Event rate per MeV per year in SK for the BH-AD diffuse neutrino background in the collapsar scenario.
Optimistic (UN) and pessimistic (0.1$\times$UN) limits as per rate estimates shown in figure \ref{ratedensity}. The SN and results 
combining the C0 model with GRBR1 rates from \cite{Nagataki:2002bn} are shown for comparison.}
\label{rate1}
\end{center}
\end{figure}

\begin{figure}[ht]
\begin{center}
\includegraphics[width=3.25 in,clip=true]
 {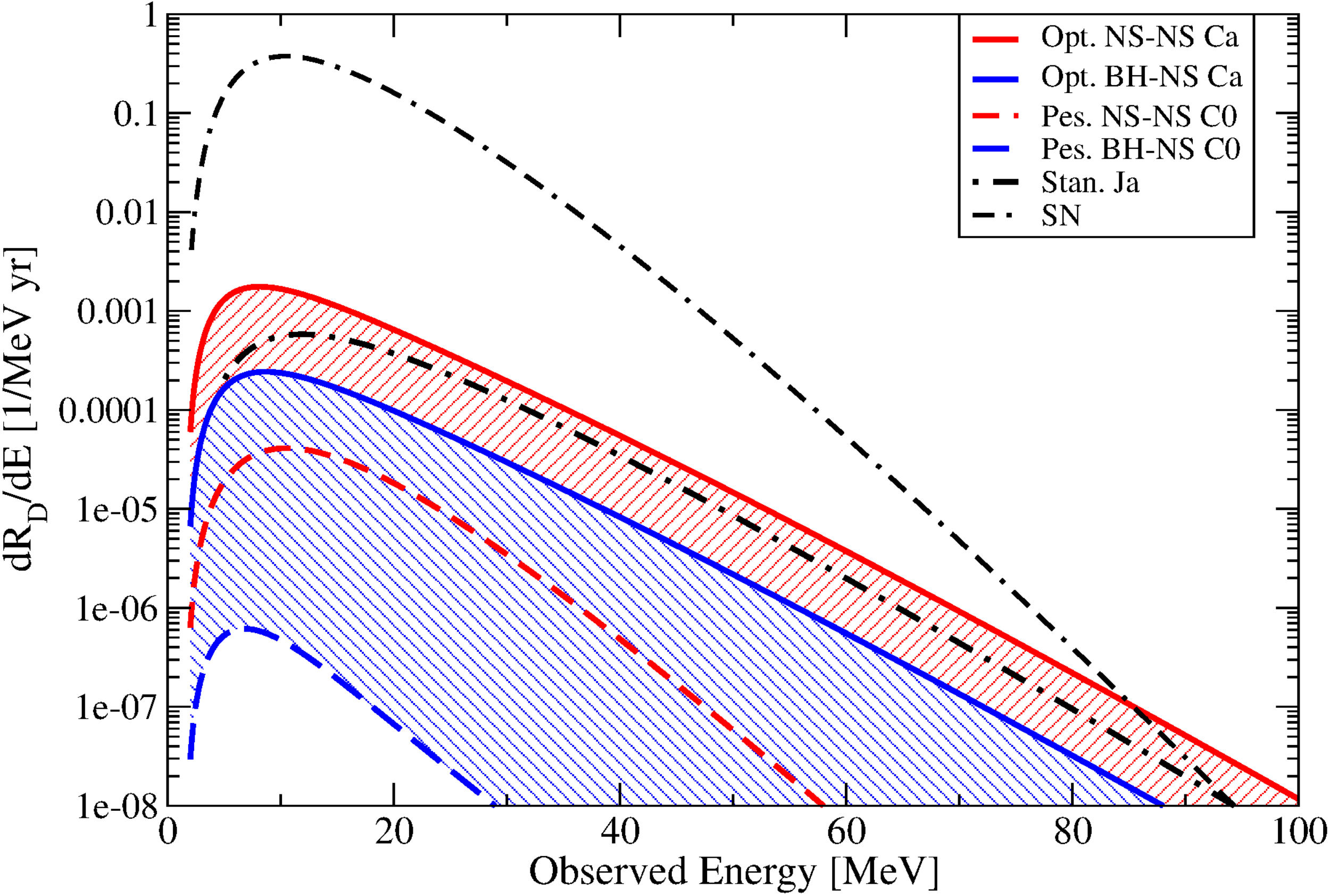}
\caption{(Color on line) Event rate per MeV per year in SK for the BH-AD diffuse neutrino background in the merger scenarios.
Optimistic and pessimistic limits as per rate estimates shown in figure \ref{ratedensity}. The SN results are shown for comparison.}
\label{rate}
\end{center}
\end{figure}

Figures \ref{rate1} and \ref{rate} show the change with energy, $dR_D/dE_o$, when the accretion disk is formed during collapsars, and BH-NS and NS-NS mergers, 
in SK assuming a 32 kton fiducial volume.
In order to study lower and upper limits of $R_D$, we have considered the most optimistic and pessimistic formation scenarios, together with
the strongest and weakest neutrino spectra. For collapsar scenarios, we take the optimistic and pessimistic collapsar rates and fold it together with a range of neutrino emission models.
These generate the extremes 
of the band shown in figure \ref{rate1}. In this way, in the collapsar scenario, the upper brown solid line corresponds to convolving the failed supernova rate with 
a electron antineutrino spectrum coming from a (Ca) disk accreting at a rate of $9 M_\odot$/sec, a BH spin $a=0.95$ and emitting neutrinos
for $\Delta t_\infty \approx 0.57$ secs; while the brown dotted line convolves a 0.1$\times$ unnova rate with a disk accreting at $3 M_\odot$/sec into a BH with a spin $a=0$ and emitting 
for 0.35 sec. For comparison we also show the detection rates found  for the latter disk model with the optimistic UN rate (thick red dashed line) and the collapsar rate, 
GRBR1, from Nagataki et al \cite{Nagataki:2002bn} (thin magenta dashed line).

We proceed in a similar fashion to evaluate the limiting lines of figure \ref{rate} in the merger scenarios. There, we have considered
the same extremes on the neutrino spectra as with the collapsar case. However, for the merger rates we use the optimistic and pessimistic occurrence rates as discussed in figure \ref{ratedensity}.
Therefore, the red solid line in figure \ref{rate} is found by multiplying the Ca spectrum ($\dot{M}=9 M_\odot$/s) with a NS-NS merger rate calculated assuming a galactic low-end metallicity evolution and 
the development of a common envelope during the compact object merger (see blue thick line in figure \ref{ratedensity}).
The pessimistic NS-NS neutrino detection rate in figure \ref{rate} (red dashed line) assumes that neutrinos have been
emitted from C0 disks with an accretion rate of $3 M_\odot$/s and occurring in galaxies with low metallicity and in the standard merger model of Dominik et al.
(green thick dotted line in figure \ref{ratedensity}). Finally, for the BH-NS merger, the optimistic neutrino detection rates (blue solid lines) correspond
to a low-end metallicity galactic evolution with common envelope 
for the BH-NS merger that evolves into the Ca disk model with $\dot{M}=9 M_\odot$/sec, while the pessimistic detection (blue dashed line) 
corresponds to a spectrum
from a C0 disk with $\dot{M}=3 M_\odot$/s and a BH-NS rate in the high-end metallicity evolution scenario with the merger producing BH kicks.
All other evolution scenarios and conditions that change the neutrino flux (e.g accretion rates), considered here, will fall 
inside these bands for the Chen-Beloborodov models. The black dot-dashed line shows the detection rate obtained for the hydrodynamical model J$a$ with
an estimated emission time of 2 sec, considering the torus was the result of a NS-NS merger (in the standard evolution scenario) happening in a galaxy with high metallicity. It is clear from the figure that if the collapsar formation rate is high, then the dominant component to the detection rates comes from collapsars, followed by a disk formed 
during a NS-NS merger and finally 
there are, due to the low occurrence rates, the BH-NS disks. 

The total number of relic neutrinos per year is obtained after integrating $dR_D/dE_o$ with energy
interval starting from the energy threshold of SK, 5 MeV, up to 100 MeV. 
The integrated rates, as per eq. \ref{detection rate} are written in Table \ref{table1}, where we also report the estimates found for the J$a$ and J$0$ models
(not shown in figure \ref{rate} for clarity).
The tabulated cases correspond to the same cases in
Figures \ref{rate1} and \ref{rate}, which, as expected from Figures \ref{spectrum}, \ref{spectrumRateSpread}, and \ref{bhspec}, shows significant increases
in detection rates stepping from BH-NS to NS-NS and to the collapsar case. We also provide re-scaled results for the 560 ktons of HyperK.

\begin{table}
\caption{Rate of relic neutrinos [1/yr] at SK (32k ton) and HK (560 kton) for the scenarios considered in figures \ref{rate1} and \ref{rate} with formation rates as in figure \ref{ratedensity}.}
\begin{center}
\begin{tabular}{c |c | c |c |c |c}
\hline
\textbf{Scenario} & \textbf{Formation} & \textbf{Disk} & $\dot{M}$ & $R_D$ SK & $R_D$ HK\\
                  & \textbf{Rate}      &    \textbf{Model}        & [$M_\odot$/s] & [1/yr] & [1/yr]\\
\hline
          &  UN        & Ca & 9 & $5.2$ &91 \\
Collapsar &  0.1xUN        & C0 & 3 & $0.06$& 1.05 \\
\hline
          & Opt.     & Ca & 9 & $2.5\times 10^{-2}$ & 0.43\\ 
NS-NS     & Pes.     & C0 & 3 & $6.0\times 10^{-4}$ &0.01\\
Merger    & Opt.     & Ja & - & $3.3\times 10^{-2}$ &0.57\\
          & Pes.     & J0 & - & $4.5\times 10^{-3}$ & 0.08\\
          & Stan.    & Ja & - & $1.0\times 10^{-2}$ & 0.17\\
\hline
          & Opt.    & Ca & 9 & $3.6\times 10^{-3}$ &$6 \times 10^{-2}$\\
BH-NS     & Pes.    & C0 & 3 & $5.4\times 10^{-6}$ &$9\times 10^{-5}$\\
Merger    & Opt.    & Ja & - & $4.7\times 10^{-3}$ &$8\times 10^{-2}$\\
          & Pes.    & J0 & - & $4.4\times 10^{-5}$ &$8\times 10^{-4}$\\
\hline
\end{tabular}
\end{center}
\label{table1}
\end{table}%

In Table \ref{table2} we summarize our detection rates for the diffuse neutrino background when
the neutrino spectra corresponds to steady-state disks (C models).
The spectra were convolved with the UN rates and with the NS-NS merger rates
in the standard coalescence scenario in galaxies with high-end metallicity.
The increased changes in detection are related to the increase in accretion rate and BH spin. 
For comparison we have included the results for the diffuse SN background found when we
take the SN rates as in eq. \ref{snrate} and consider a SN spectrum as in ref. \cite{Totani:1997vj}.
We show neutrino counts for two energy windows, one above SuperK threshold, from 5 MeV to 100 MeV, and another one
that corresponds to the current allowed interval, above detector backgrounds (atmospheric and reactor backgrounds), from 11 to 30 MeV.

\begin{table}
\caption{Number of relic neutrinos per year, $R_D$, from collapsars and 
NS-NS mergers, assuming the remnant disks accrete with a fixed rate $\dot{M}$  and
BH spin $a$. Rates are given for two sets of energy windows in Super-K. Results for SN are provided for comparison.}
\begin{center}
\begin{tabular}{c |cc |cc}
\hline
\multicolumn{5}{c}{$E_\nu>5$MeV}  \\
 \hline
$R_D$ [1/yr] & \multicolumn {2}{c|}{Collapsar} &\multicolumn {2}{c} {NS-NS} ($\times 10^{-3}$)\\ 
\hline
 $\dot{M}$  &$a = 0$&$a = 0.95$ &$a = 0$&$a = 0.95$ \\
 \hline

$3 M_\odot/s$&0.5& 2.3& $0.9$& $3.4$ \\
$5 M_\odot/s$&0.8 & 3.4&$1.4$& $5.3$\\
$7 M_\odot/s$&1.0 &4.4 & $1.7$&$6.8$\\
$9 M_\odot/s$&1.3& 5.2 & $2.1$&$8 $ \\
\hline
SN        & \multicolumn {2}{c|}{5.3}&\\
\hline
\hline
\multicolumn{5}{c}{$11<E_\nu <30$MeV}  \\
 \hline
$3 M_\odot/s$&0.2&1.2 &$0.51$&2.2 \\

$5 M_\odot/s$&0.3&1.8 &$0.77$&3.3 \\
$7 M_\odot/s$&0.4&2.3 & $1.0$&4.1\\
$9 M_\odot/s$&0.5&2.6 &$1.1$&$4.9$\\
\hline
SN        & \multicolumn {2}{c|}{3.3}&\\
\hline
\end{tabular}
\end{center}
\label{table2}
\end{table}%

%%%%%%%%%%%%%%%%%%%%%%%%%%%%%%%%%%%%%%%%%%%%%%%%%%%%%%%%%%%%%%%%%%%%%%%%%%%%%%%
\section{Summary and Conclusions}
\label{conclusions}
%%%%%%%%%%%%%%%%%%%%%%%%%%%%%%%%%%%%%%%%%%%%%%%%%%%%%%%%%%%%%%%%%%%%%%%%%%%%%%%
We have studied the spectra, diffuse fluxes, and detection rates in SuperK and HyperK, of neutrinos
emitted by past to present black hole accretion disks. Those are considered to be one of the possible 
final fates of rotating core collapse supernovae, as well as of  mergers of  neutron stars with  black holes or with other neutron stars. The models used
for our study include important aspects such neutrino cooling and relativistic effects.

Neutrino
disk spectra depend on the mass accretion rate and the BH spin.
The evolution of accretion disks is such that there is a funnel formed around the black hole. When neutrinos are emitted from that low density region they
have large temperatures. The effect of these high temperatures and the torus like geometry is reflected in a hotter neutrino spectrum compared to 
that from spherically symmetric SN simulations (the latter used to study the failed SN spectrum).
The number of failed supernovae that evolve into a disk (in a collapsar model) depends on
still to be determined
physics such as the nuclear matter equation of state and the progenitor initial conditions, leaving us with open questions on
the mechanism of BH formation. Future simulations would shed light into
the BH mass, BH spin, and accretion rate ranges that would be possible if such tori formed from unsuccessful SN. 

This uncertainty notwithstanding, our spectra results motivated us to study the potential contribution of these neutrinos to
the relic neutrino background in the MeV range. We find that in the collapsar model, assuming an upper limit event rate that is the same as the unnova rate, 
this diffuse flux is comparable (larger for high mass accretion rates) to the SN one.
We find that the number of neutrinos registered in SuperK (taking an energy threshold of 5 MeV) in a 5 year period from collapsars would be between 3 to 25.
As discussed elsewhere (see e. g. \cite{Priya:2017bmm}), the atmospheric and reactor neutrino fluxes limit the detection energy window from $\sim$ 11 to 30 MeV in the
current SuperK setup. In that range we predict that in the most optimistic collapsar model we will find about 3 counts per year. 

We also studied the diffuse flux and possible detection of neutrinos coming from ADs in the scenario where the torus was
the result of a neutron star-compact object merger. The cosmological merger rates lead to diffuse fluxes that are at least two orders of magnitude
smaller than those of SN and collapsars. However, the upgrade from SuperK to HyperK will allow for a detection of 
at least one of those neutrinos (in the most optimistic merger scenario) in a period of 1.75 years. 
It is important also to keep in mind that these results are based on
merger rates for field stellar populations \cite{dominik}, but rates should be larger 
in globular clusters. The rates are also sensitive to parameters in the binary model
and initial distributions of the binary \cite{deMink:2015yea}. A recent compilation of different 
predictions of NS-NS and BH-NS merger rates can be found in \cite{Abbott:2016ymx} showing that event rates may be higher
than assumed here.

The prospects of overcoming the current detection limitations on the detection of the CMNB are promising. Extracting relic neutrino signals
in SK, with more data collected, improved efficiency, and lower threshold will be a reality in few years \cite{Zhang:2013tua}. 
The possibility of a megaton water-Cherenkov detector like HyperK opens the door to significant numbers of diffuse neutrinos being observed. In analyzing such a future detection, we should bear in mind that in addition to standard core collapse and failed supernovae, a few of these neutrinos may come from accretion disk supernovae and compact object mergers.

%%%%%%%%%%%%%%%%%%%%%%%%%%%%%%%%%%%%%%%%%%%%%%%%%%%%%%%%%%%%%%%%%%%%%%%%%%%%%%%  
\section{Acknowledgments}
%%%%%%%%%%%%%%%%%%%%%%%%%%%%%%%%%%%%%%%%%%%%%%%%%%%%%%%%%%%%%%%%%%%%%%%%%%%%%%%

OLC thanks Eric Poisson for useful discussions. This work was partially supported by the Natural Sciences and
Engineering Research Council of Canada (NSERC)(OLC), the U.S. DOE Grant
No. DE-FG02-02ER41216 and the National Science Foundation, Grant PHY-1630782(GCM). 
%and DE-SC0004786(GCM). 

%%%%%%%%%%%%%%%%%%%%%%%%%%%%%%%%%%%%%%%%%%%%%%%%%%%%%%%%%%%%%%%%%%%%%%%%%%%%%%%

%%%%%%%%%%%%%%%%%%%%%%%%%%%%%%%%%%%%%%%%%%%%%%%%%%%%%%%%%%%%%%%%%%%%%%%%%%%%%%%
\vfill\eject


\begin{thebibliography}{99}
%%%%%%%%%%%%%%%%%%%%%%%%%%%%%%%%%%%%%%%%%%%%%%%%%%%%%%%%%%%%%%%%%%%%%%%%%%%%%%%


\bibitem{Lee1999} W. Kluzniak and W. H. Lee,  MNRAS {\bf 308} (1999) 780.



\bibitem{Rosswog:2005su} 
  S.~Rosswog,
  %``Mergers of neutron star black hole binaries with small mass ratios: Nucleosynthesis, gamma-ray bursts and electromagnetic transients,''
  Astrophys.\ J.\  {\bf 634}, 1202 (2005)
%  doi:10.1086/497062
%  [astro-ph/0508138].

%\cite{Fryer:2015uia}
\bibitem{Fryer:2015uia} 
  C.~L.~Fryer, K.~Belczynski, E.~Ramirez-Ruiz, S.~Rosswog, G.~Shen and A.~W.~Steiner,
  %``The Fate of the Compact Remnant in Neutron Star Mergers,''
  Astrophys.\ J.\  {\bf 812}, no. 1, 24 (2015)
%  doi:10.1088/0004-637X/812/1/24

%\cite{Foucart:2014nda}
\bibitem{Foucart:2014nda} 
  F.~Foucart {\it et al.},
  %``Neutron star-black hole mergers with a nuclear equation of state and neutrino cooling: Dependence in the binary parameters,''
  Phys.\ Rev.\ D {\bf 90}, 024026 (2014)
%  doi:10.1103/PhysRevD.90.024026
%  [arXiv:1405.1121 [astro-ph.HE]].
  
  %\cite{Lehner:2016lxy}
\bibitem{Lehner:2016lxy} 
  L.~Lehner, S.~L.~Liebling, C.~Palenzuela, O.~L.~Caballero, E.~O'Connor, M.~Anderson and D.~Neilsen,
  %``Unequal mass binary neutron star mergers and multimessenger signals,''
  Class.\ Quant.\ Grav.\  {\bf 33}, no. 18, 184002 (2016)
%  doi:10.1088/0264-9381/33/18/184002
%  [arXiv:1603.00501 [gr-qc]].
  %%CITATION = doi:10.1088/0264-9381/33/18/184002;%%

%\cite{MacFadyen:1998vz}
\bibitem{MacFadyen:1998vz} 
  A.~MacFadyen and S.~E.~Woosley,
  %``Collapsars: Gamma-ray bursts and explosions in 'failed supernovae',''
  Astrophys.\ J.\  {\bf 524}, 262 (1999)
%  doi:10.1086/307790
%  [astro-ph/9810274].
  %%CITATION = doi:10.1086/307790;%%

%\cite{OConnor:2010moj}
\bibitem{OConnor:2010moj} 
  E.~O'Connor and C.~D.~Ott,
  %``Black Hole Formation in Failing Core-Collapse Supernovae,''
  Astrophys.\ J.\  {\bf 730}, 70 (2011)
%  doi:10.1088/0004-637X/730/2/70
%  [arXiv:1010.5550 [astro-ph.HE]].
  %%CITATION = doi:10.1088/0004-637X/730/2/70;%%
  
%\cite{Ott:2010gv}
\bibitem{Ott:2010gv} 
  C.~D.~Ott {\it et al.},
  %``Dynamics and Gravitational Wave Signature of Collapsar Formation,''
  Phys.\ Rev.\ Lett.\  {\bf 106}, 161103 (2011)
%  doi:10.1103/PhysRevLett.106.161103
%  [arXiv:1012.1853 [astro-ph.HE]].
  %%CITATION = doi:10.1103/PhysRevLett.106.161103;%%
  
  %\cite{Sekiguchi:2010ja}
\bibitem{Sekiguchi:2010ja} 
  Y.~Sekiguchi and M.~Shibata,
  %``Formation of black hole and accretion disk in a massive high-entropy stellar core collapse,''
  Astrophys.\ J.\  {\bf 737}, 6 (2011)
%  doi:10.1088/0004-637X/737/1/6
%  [arXiv:1009.5303 [astro-ph.HE]].
  
 \bibitem{Surman_Mclaughlin04} R. Surman and G. C. McLaughlin, ApJ {\bf 603} (2004) 611. 
\bibitem{Surmanrprocess} R. Surman and G. C. McLaughlin, ApJ, {\bf 679} (2008) L117.
  
  %\cite{Just:2014fka}
\bibitem{Just:2014fka} 
  O.~Just, A.~Bauswein, R.~A.~Pulpillo, S.~Goriely and H.-T.~Janka,
  %``Comprehensive nucleosynthesis analysis for ejecta of compact binary mergers,''
  Mon.\ Not.\ Roy.\ Astron.\ Soc.\  {\bf 448}, no. 1, 541 (2015)
%  doi:10.1093/mnras/stv009
%  [arXiv:1406.2687 [astro-ph.SR]].
  %%CITATION = doi:10.1093/mnras/stv009;%%
  
  %\cite{Caballero:2011dw}
\bibitem{Caballero:2011dw} 
  O.~L.~Caballero, G.~C.~McLaughlin and R.~Surman,
  %``Neutrino Spectra from Accretion Disks: Neutrino General Relativistic Effects and the Consequences for Nucleosynthesis,''
  Astrophys.\ J.\  {\bf 745}, 170 (2012)
%  doi:10.1088/0004-637X/745/2/170
%  [arXiv:1105.6371 [astro-ph.HE]].


%\cite{Surman:2008qf}
\bibitem{Surman:2008qf} 
  R.~Surman, G.~C.~McLaughlin, M.~Ruffert, H.-T.~Janka and W.~R.~Hix,
  %``r-Process Nucleosynthesis in Hot Accretion Disk Flows from Black Hole - Neutron Star Mergers,''
  Astrophys.\ J.\  {\bf 679}, L117 (2008)
  
  %\cite{Woosley:1993wj}
\bibitem{Woosley:1993wj} 
  S.~E.~Woosley,
  %``Gamma-ray bursts from stellar mass accretion disks around black holes,''
  Astrophys.\ J.\  {\bf 405}, 273 (1993).
%  doi:10.1086/172359
  %%CITATION = doi:10.1086/172359;%%
  
  \bibitem{Popham1999} Robert Popham, S. E. Woosley and Chris Fryer, ApJ {\bf 518} (1999) 356.
  
  %\cite{Murguia-Berthier:2016fys}
\bibitem{Murguia-Berthier:2016fys} 
  A.~Murguia-Berthier {\it et al.},
  %``The Properties of Short gamma-ray burst Jets Triggered by neutron star mergers,''
  Astrophys.\ J.\  {\bf 835}, no. 2, L34 (2017)
%  doi:10.3847/2041-8213/aa5b9e
%  [arXiv:1609.04828 [astro-ph.HE]].

%\cite{Kneller:2004jr}
\bibitem{Kneller:2004jr} 
  J.~P.~Kneller, G.~C.~McLaughlin and R.~Surman,
  %``Neutrino scattering, absorption and annihilation above the accretion disks of gamma ray bursts,''
  J.\ Phys.\ G {\bf 32}, 443 (2006)


  
  %\cite{Hirata:1987hu}
\bibitem{Hirata:1987hu} 
  K.~Hirata {\it et al.} [Kamiokande-II Collaboration],
  %``Observation of a Neutrino Burst from the Supernova SN 1987a,''
  Phys.\ Rev.\ Lett.\  {\bf 58}, 1490 (1987).
%  doi:10.1103/PhysRevLett.58.1490

 \bibitem{SK} S. Fukuda et al., Nucl. Inst. Meth. Phys. A, {\bf 501} (2003) 418.
 
 %\cite{Descamps:2015hva}
\bibitem{Descamps:2015hva} 
  F.~Descamps [SNO+ Collaboration],
  %``Neutrino Physics with SNO+,''
  Nucl.\ Part.\ Phys.\ Proc.\  {\bf 265-266}, 143 (2015).
%  doi:10.1016/j.nuclphysbps.2015.06.037



\bibitem{hk}Kenzo Nakamura, Int. J. Mod. Phys. A, {\bf 18}(2003)4053.


\bibitem{uno}C. K. Jung International workshop on Next Generation Nucleon Decay and Neutrino Detectors AIP conference proccedings {\bf 533}. Arxiv:hep-ex/0005046.

%\cite{Acciarri:2015uup}
\bibitem{Acciarri:2015uup} 
  R.~Acciarri {\it et al.} [DUNE Collaboration],
  %``Long-Baseline Neutrino Facility (LBNF) and Deep Underground Neutrino Experiment (DUNE) : Volume 2: The Physics Program for DUNE at LBNF,''
  arXiv:1512.06148 [physics.ins-det].
  
%\cite{An:2015jdp}
\bibitem{An:2015jdp} 
  F.~An {\it et al.} [JUNO Collaboration],
  %``Neutrino Physics with JUNO,''
  J.\ Phys.\ G {\bf 43}, no. 3, 030401 (2016)  
  
\bibitem{Titand}Y. Suzuki, hep-ex/0110005.

  %\cite{Kistler:2008us}
\bibitem{Kistler:2008us} 
  M.~D.~Kistler, H.~Yuksel, S.~Ando, J.~F.~Beacom and Y.~Suzuki,
  %``Core-Collapse Astrophysics with a Five-Megaton Neutrino Detector,''
  Phys.\ Rev.\ D {\bf 83}, 123008 (2011)
%  doi:10.1103/PhysRevD.83.123008
%  [arXiv:0810.1959 [astro-ph]].
  
  %\cite{Nakazato:2015rya}
\bibitem{Nakazato:2015rya} 
  K.~Nakazato, E.~Mochida, Y.~Niino and H.~Suzuki,
  %``Spectrum of the Supernova Relic Neutrino Background and Metallicity Evolution of Galaxies,''
  Astrophys.\ J.\  {\bf 804}, no. 1, 75 (2015)
%  doi:10.1088/0004-637X/804/1/75
%  [arXiv:1503.01236 [astro-ph.HE]].
  %%CITATION = doi:10.1088/0004-637X/804/1/75;%%

 %\cite{Davis:2017mbq}
\bibitem{Davis:2017mbq} 
  J.~H.~Davis and M.~Fairbairn,
  %``A "nu" look at gravitational waves: The black hole birth rate from neutrinos combined with the merger rate from LIGO,''
  JCAP {\bf 1707}, no. 07, 052 (2017)
%  doi:10.1088/1475-7516/2017/07/052
%  [arXiv:1704.05073 [astro-ph.HE]]. 

% 
% %\cite{GBM:2017lvd}
% \bibitem{GBM:2017lvd} 
%   B.~P.~Abbott {\it et al.} [LIGO Scientific and Virgo and Fermi GBM and INTEGRAL and IceCube and IPN and Insight-Hxmt and ANTARES and Swift and Dark Energy Camera GW-EM and Dark Energy Survey and DLT40 and GRAWITA and Fermi-LAT and ATCA and ASKAP and OzGrav and DWF (Deeper Wider Faster Program) and AST3 and CAASTRO and VINROUGE and MASTER and J-GEM and GROWTH and JAGWAR and CaltechNRAO and TTU-NRAO and NuSTAR and Pan-STARRS and KU and Nordic Optical Telescope and ePESSTO and GROND and Texas Tech University and TOROS and BOOTES and MWA and CALET and IKI-GW Follow-up and H.E.S.S. and LOFAR and LWA and HAWC and Pierre Auger and ALMA and Pi of Sky and DFN and ATLAS Telescopes and High Time Resolution Universe Survey and RIMAS and RATIR and SKA South Africa/MeerKAT Collaborations and AstroSat Cadmium Zinc Telluride Imager Team and AGILE Team and 1M2H Team and Las Cumbres Observatory Group and MAXI Team and TZAC Consortium and SALT Group and Euro VLBI Team and Chandra Team at McGill University],
%   %``Multi-messenger Observations of a Binary Neutron Star Merger,''
%   Astrophys.\ J.\  {\bf 848}, no. 2, L12 (2017)
 
  
%\cite{Lunardini:2010ab}
\bibitem{Lunardini:2010ab} 
  C.~Lunardini,
  %``Diffuse supernova neutrinos at underground laboratories,''
  Astropart.\ Phys.\  {\bf 79}, 49 (2016)
%  doi:10.1016/j.astropartphys.2016.02.005
%  [arXiv:1007.3252 [astro-ph.CO]].
  %%CITATION = doi:10.1016/j.astropartphys.2016.02.005;%%
  %34 citations counted in INSPIRE as of 17 Aug 2017
  



\bibitem{Ando2003}S. Ando, K. Sato and T. Totani, Astroparticle Physics {\bf 18} (2003) 307.

\bibitem{Ando2004} 
  S.~Ando and K.~Sato,
  %``Relic neutrino background from cosmological supernovae,''
  New J.\ Phys.\  {\bf 6}, 170 (2004)
%  doi:10.1088/1367-2630/6/1/170
%  [astro-ph/0410061].


\bibitem{Lunardini2006} Cecilia Lunardini, Phys.Rev. {\bf D73} (2006) 083009  


\bibitem{Strigari}Strigari, L. et al, JCAP 0403 (2004) 007.

%\cite{Kaplinghat:1999xi}
\bibitem{Kaplinghat:1999xi} 
  M.~Kaplinghat, G.~Steigman and T.~P.~Walker,
  %``The Supernova relic neutrino background,''
  Phys.\ Rev.\ D {\bf 62}, 043001 (2000)
%  doi:10.1103/PhysRevD.62.043001
%  [astro-ph/9912391].
  %%CITATION = doi:10.1103/PhysRevD.62.043001;%%
  
  
  %\cite{Beacom:2003nk}
\bibitem{Beacom:2003nk} 
  J.~F.~Beacom and M.~R.~Vagins,
  %``GADZOOKS! Anti-neutrino spectroscopy with large water Cherenkov detectors,''
  Phys.\ Rev.\ Lett.\  {\bf 93}, 171101 (2004)
  doi:10.1103/PhysRevLett.93.171101

     
  %\cite{Beacom:2010kk}
\bibitem{Beacom:2010kk} 
  J.~F.~Beacom,
  %``The Diffuse Supernova Neutrino Background,''
  Ann.\ Rev.\ Nucl.\ Part.\ Sci.\  {\bf 60}, 439 (2010)
%  doi:10.1146/annurev.nucl.010909.083331
%  [arXiv:1004.3311 [astro-ph.HE]].

%\cite{Malek:2002ns}
\bibitem{Malek:2002ns} 
  M.~Malek {\it et al.} [Super-Kamiokande Collaboration],
  %``Search for supernova relic neutrinos at SUPER-KAMIOKANDE,''
  Phys.\ Rev.\ Lett.\  {\bf 90}, 061101 (2003)

%\cite{Bays:2011si}
\bibitem{Bays:2011si} 
  K.~Bays {\it et al.} [Super-Kamiokande Collaboration],
  %``Supernova Relic Neutrino Search at Super-Kamiokande,''
  Phys.\ Rev.\ D {\bf 85}, 052007 (2012)

%\cite{Zhang:2013tua}
\bibitem{Zhang:2013tua} 
  H.~Zhang {\it et al.} [Super-Kamiokande Collaboration],
  %``Supernova Relic Neutrino Search with Neutron Tagging at Super-Kamiokande-IV,''
  Astropart.\ Phys.\  {\bf 60}, 41 (2015)



 

\bibitem{Priya:2017bmm} 
  A.~Priya and C.~Lunardini,
  %``Diffuse neutrinos from luminous and dark supernovae: prospects for upcoming detectors at the O(10) kt scale,''
  arXiv:1705.02122 [astro-ph.HE].
  %%CITATION = ARXIV:1705.02122;%%
  
  %\cite{Yang:2011xd}
\bibitem{Yang:2011xd} 
  L.~Yang and C.~Lunardini,
  %``Revealing local failed supernovae with neutrino telescopes,''
  Phys.\ Rev.\ D {\bf 84}, 063002 (2011)
%  doi:10.1103/PhysRevD.84.063002
%  [arXiv:1103.4628 [astro-ph.CO]].
  %%CITATION = doi:10.1103/PhysRevD.84.063002;%%
  %10 citations counted in INSPIRE as of 17 Aug 2017
  
    %\cite{Keehn:2010pn}
\bibitem{Keehn:2010pn} 
  J.~G.~Keehn and C.~Lunardini,
  %``Neutrinos from failed supernovae at future water and liquid argon detectors,''
  Phys.\ Rev.\ D {\bf 85}, 043011 (2012)
  
\bibitem{Lunardini2009failed} Cecilia Lunardini, Phys. Rev. Lett. {\bf 102} (2009) 231101.

  
%\cite{Nakazato:2008vj}
\bibitem{Nakazato:2008vj} 
  K.~Nakazato, K.~Sumiyoshi, H.~Suzuki and S.~Yamada,
  %``Oscillation and Future Detection of Failed Supernova Neutrinos from Black Hole Forming Collapse,''
  Phys.\ Rev.\ D {\bf 78}, 083014 (2008)
  Erratum: [Phys.\ Rev.\ D {\bf 79}, 069901 (2009)]
%  doi:10.1103/PhysRevD.78.083014, 10.1103/PhysRevD.79.069901
%  [arXiv:0810.3734 [astro-ph]].
  %%CITATION = doi:10.1103/PhysRevD.78.083014, 10.1103/PhysRevD.79.069901;%%
  %30 citations counted in INSPIRE as of 17 Aug 2017

%\cite{Nagataki:2002bn}
\bibitem{Nagataki:2002bn} 
  S.~Nagataki, K.~Kohri, S.~Ando and K.~Sato,
  %``Gamma-ray burst neutrino background and star formation history in the universe,''
  Astropart.\ Phys.\  {\bf 18}, 551 (2003)  
%\bibitem{Nagataki2003}S. Nagataki, K. Kohri, S. Ando and Katsuhiko Sato, Nuclear Physics {\bf A718} (2003) 473c.
\bibitem{Nagatakicounts} Shigehiro Nagataki and Kazunori Kohri, Prog. Theor. Phys., {\bf 108} (2002)789.
  
%\cite{Fischer:2008rh}
\bibitem{Fischer:2008rh} 
  T.~Fischer, S.~C.~Whitehouse, A.~Mezzacappa, F.-K.~Thielemann and M.~Liebendorfer,
  %``The neutrino signal from protoneutron star accretion and black hole formation,''
  Astron.\ Astrophys.\  {\bf 499}, 1 (2009)
%  doi:10.1051/0004-6361/200811055
%  [arXiv:0809.5129 [astro-ph]].  
  
  %\cite{Sumiyoshi:2008zw}
\bibitem{Sumiyoshi:2008zw} 
  K.~Sumiyoshi, S.~Yamada and H.~Suzuki,
  %``Dynamics and neutrino signal of black hole formation in non-rotating failed supernovae. II. progenitor dependence,''
  Astrophys.\ J.\  {\bf 688}, 1176 (2008)
%  doi:10.1086/592183
%  [arXiv:0808.0384 [astro-ph]]

\bibitem{Cab2009} 
  O.~L.~Caballero, G.~C.~McLaughlin, R.~Surman and R.~Surman,
  %``Detecting neutrinos from black hole neutron stars mergers,''
  Phys.\ Rev.\ D {\bf 80}, 123004 (2009)
%  doi:10.1103/PhysRevD.80.123004
%  [arXiv:0910.1385 [astro-ph.HE]].
  
  
  %\cite{OShaughnessy:2008yul}
\bibitem{OShaughnessy:2008yul} 
  R.~W.~O'Shaughnessy, R.~K.~Kopparapu and K.~Belczynski,
  %``Impact of star formation inhomogeneities on merger rates and interpretation of LIGO results,''
  Class.\ Quant.\ Grav.\  {\bf 29}, 145011 (2012)
%  doi:10.1088/0264-9381/29/14/145011
%  [arXiv:0812.0591 [astro-ph]].

\bibitem{dominik} 
  M.~Dominik, K.~Belczynski, C.~Fryer, D.~E.~Holz, E.~Berti, T.~Bulik, I.~Mandel and R.~O'Shaughnessy,
  %``Double Compact Objects II: Cosmological Merger Rates,''
  Astrophys.\ J.\  {\bf 779}, 72 (2013)
%  doi:10.1088/0004-637X/779/1/72
%  [arXiv:1308.1546 [astro-ph.HE]].


  \bibitem{Abadie:2010cf} 
  J.~Abadie {\it et al.} [LIGO Scientific and VIRGO Collaborations],
  %``Predictions for the Rates of Compact Binary Coalescences Observable by Ground-based Gravitational-wave Detectors,''
  Class.\ Quant.\ Grav.\  {\bf 27}, 173001 (2010)
% 
% %\cite{Dominik:2012kk}
% \bibitem{Dominik:2012kk} 
%   M.~Dominik, K.~Belczynski, C.~Fryer, D.~Holz, E.~Berti, T.~Bulik, I.~Mandel and R.~O'Shaughnessy,
%   %``Double Compact Objects I: The Significance of the Common Envelope on Merger Rates,''
%   Astrophys.\ J.\  {\bf 759}, 52 (2012)
%   doi:10.1088/0004-637X/759/1/52
%   [arXiv:1202.4901 [astro-ph.HE]].
%   
  
%\cite{TheLIGOScientific:2017qsa}
\bibitem{TheLIGOScientific:2017qsa} 
  B.~P.~Abbott {\it et al.} [LIGO Scientific and Virgo Collaborations],
  %``GW170817: Observation of Gravitational Waves from a Binary Neutron Star Inspiral,''
  Phys.\ Rev.\ Lett.\  {\bf 119}, no. 16, 161101 (2017)
  
% %\cite{Fryer:2011cx}
% \bibitem{Fryer:2011cx} 
%   C.~L.~Fryer, K.~Belczynski, G.~Wiktorowicz, M.~Dominik, V.~Kalogera and D.~E.~Holz,
%   %``Compact Remnant Mass Function: Dependence on the Explosion Mechanism and Metallicity,''
%   Astrophys.\ J.\  {\bf 749}, 91 (2012)
%   doi:10.1088/0004-637X/749/1/91
%   [arXiv:1110.1726 [astro-ph.SR]].



%\cite{Gehrels:2004aa}
\bibitem{Gehrels:2004aa} 
  N.~Gehrels {\it et al.} [Swift Science Collaboration],
  %``The Swift Gamma-Ray Burst Mission,''
  Astrophys.\ J.\  {\bf 611}, 1005 (2004)
  Erratum: [Astrophys.\ J.\  {\bf 621}, 558 (2005)]
  
  %\cite{Butler:2007hw}
\bibitem{Butler:2007hw} 
  N.~R.~Butler, D.~Kocevski, J.~S.~Bloom and J.~L.~Curtis,
  %``A Complete Catalog of Swift GRB Spectra and Durations: Demise of a Physical Origin for Pre-Swift High-Energy Correlations,''
  Astrophys.\ J.\  {\bf 671}, 656 (2007)
  
  \bibitem{Yuksel2008} 
  H.~Yuksel, M.~D.~Kistler, J.~F.~Beacom and A.~M.~Hopkins,
  %``Revealing the High-Redshift Star Formation Rate with Gamma-Ray Bursts,''
  Astrophys.\ J.\  {\bf 683}, L5 (2008)
%  doi:10.1086/591449
%  [arXiv:0804.4008 [astro-ph]].
  
\bibitem{Yuksel:2012zy} 
  H.Yüksel and M.D. Kistler,
  %``The cosmic MeV neutrino background as a laboratory for black hole formation,''
  Phys.\ Lett.\ B {\bf 751}, 413 (2015)
%  doi:10.1016/j.physletb.2015.10.055
%  [arXiv:1212.4844 [astro-ph.HE]]. 

%\cite{Lunardini:2012ne}
\bibitem{Lunardini:2012ne} 
  C.~Lunardini and I.~Tamborra,
  %``Diffuse supernova neutrinos: oscillation effects, stellar cooling and progenitor mass dependence,''
  JCAP {\bf 1207}, 012 (2012)
%  doi:10.1088/1475-7516/2012/07/012
%  [arXiv:1205.6292 [astro-ph.SR]].
  %%CITATION = doi:10.1088/1475-7516/2012/07/012;%%
  %16 citations counted in INSPIRE as of 15 Sep 2017  

%\cite{Zhu:2016mwa}
\bibitem{Zhu:2016mwa} 
  Y.~L.~Zhu, A.~Perego and G.~C.~McLaughlin,
  %``Matter Neutrino Resonance Transitions above a Neutron Star Merger Remnant,''
  Phys.\ Rev.\ D {\bf 94}, no. 10, 105006 (2016)
  doi:10.1103/PhysRevD.94.105006
  [arXiv:1607.04671 [hep-ph]].
  %%CITATION = doi:10.1103/PhysRevD.94.105006;%%
  %19 citations counted in INSPIRE as of 16 May 2018

%\cite{Malkus:2014iqa}
\bibitem{Malkus:2014iqa} 
  A.~Malkus, A.~Friedland and G.~C.~McLaughlin,
  %``Matter-Neutrino Resonance Above Merging Compact Objects,''
  arXiv:1403.5797 [hep-ph].
  %%CITATION = ARXIV:1403.5797;%%
  %30 citations counted in INSPIRE as of 16 May 2018

%\cite{Malkus:2012ts}
\bibitem{Malkus:2012ts} 
  A.~Malkus, J.~P.~Kneller, G.~C.~McLaughlin and R.~Surman,
  %``Neutrino oscillations above black hole accretion disks: disks with electron-flavor emission,''
  Phys.\ Rev.\ D {\bf 86}, 085015 (2012)
  doi:10.1103/PhysRevD.86.085015
  [arXiv:1207.6648 [hep-ph]].
  %%CITATION = doi:10.1103/PhysRevD.86.085015;%%
  %38 citations counted in INSPIRE as of 16 May 2018

  
 
\bibitem{Setiawan04} S. Setiawan, M. Ruffert and H.-Th. Janka, Mon. Not. R. Astron. Soc., {\bf 352} (2004) 753.    




%\cite{Perego:2014fma}
\bibitem{Perego:2014fma} 
  A.~Perego, S.~Rosswog, R.~M.~Cabezón, O.~Korobkin, R.~Käppeli, A.~Arcones and M.~Liebendörfer,
  %``Neutrino-driven winds from neutron star merger remnants,''
  Mon.\ Not.\ Roy.\ Astron.\ Soc.\  {\bf 443}, no. 4, 3134 (2014)
  doi:10.1093/mnras/stu1352

  
\bibitem{Beloborodovcross}Wen-Xin Chen and Andrei M. Beloborodov, ApJ {\bf657} (2007) 383.


%\bibitem{SetiawanBHmerger} S. Setiawan, M. Ruffert and H.-Th. Janka, Astron. Astrophys., {\bf 458} (2006), 553.
%\bibitem{Matteo02} Tiziana Di Matteo, Rosalba Perna and Ramesh Narayan, ApJ {\bf 579} (2002) 706.





%\cite{Shakura:1972te}
\bibitem{Shakura:1972te} 
  N.~I.~Shakura and R.~A.~Sunyaev,
  %``Black holes in binary systems. Observational appearance,''
  Astron.\ Astrophys.\  {\bf 24}, 337 (1973).
  
 %\cite{Shapiro:2017cny}
\bibitem{Shapiro:2017cny} 
  S.~L.~Shapiro,
  %``Black holes, disks, and jets following binary mergers and stellar collapse: The narrow range of electromagnetic luminosities and accretion rates,''
  Phys.\ Rev.\ D {\bf 95}, no. 10, 101303 (2017)
%  doi:10.1103/PhysRevD.95.101303
%  [arXiv:1705.04695 [astro-ph.HE]]. 
  
  
%\cite{Shibata:2006nm}
\bibitem{Shibata:2006nm} 
  M.~Shibata and K.~Taniguchi,
  %``Merger of binary neutron stars to a black hole: disk mass, short gamma-ray bursts, and quasinormal mode ringing,''
  Phys.\ Rev.\ D {\bf 73}, 064027 (2006)
%  doi:10.1103/PhysRevD.73.064027
%  [astro-ph/0603145].
  
 %\cite{Shibata:2005mz}
\bibitem{Shibata:2005mz} 
  M.~Shibata, M.~D.~Duez, Y.~T.~Liu, S.~L.~Shapiro and B.~C.~Stephens,
  %``Magnetized hypermassive neutron star collapse: A Central engine for short gamma-ray bursts,''
  Phys.\ Rev.\ Lett.\  {\bf 96}, 031102 (2006)
%  doi:10.1103/PhysRevLett.96.031102
%  [astro-ph/0511142]. 
  
 %\cite{Nouri:2017fvh}
\bibitem{Nouri:2017fvh} 
  F.~H.~Nouri {\it et al.},
  %``Evolution of the Magnetized, Neutrino-Cooled Accretion Disk in the Aftermath of a Black Hole Neutron Star Binary Merger,''
  arXiv:1710.07423 [astro-ph.HE]. 
  
 %\cite{Fernandez:2013tya}
\bibitem{Fernandez:2013tya} 
  R.~Fernández and B.~D.~Metzger,
  %``Delayed outflows from black hole accretion tori following neutron star binary coalescence,''
  Mon.\ Not.\ Roy.\ Astron.\ Soc.\  {\bf 435}, 502 (2013)
%  doi:10.1093/mnras/stt1312
%  [arXiv:1304.6720 [astro-ph.HE]].

  
   \bibitem{Deaton:2013sla} 
  M.~B.~Deaton {\it et al.},
  %``Black Hole-Neutron Star Mergers with a Hot Nuclear Equation of State: Outflow and Neutrino-Cooled Disk for a Low-Mass, High-Spin Case,''
  Astrophys.\ J.\  {\bf 776}, 47 (2013) 
  
  \bibitem{Artemova} Artemova I. V., Bjoernsson G., Novikov I. D., 1996, ApJ,
461, 565

\bibitem{Caballero:2015cpa} 
  O.~L.~Caballero, T.~Zielinski, G.~C.~McLaughlin and R.~Surman,
  %``Black hole spin influence on accretion disk neutrino detection,''
  Phys.\ Rev.\ D {\bf 93}, no. 12, 123015 (2016)
  
 
%\cite{Hotokezaka:2013iia}
\bibitem{Hotokezaka:2013iia} 
  K.~Hotokezaka, K.~Kiuchi, K.~Kyutoku, T.~Muranushi, Y.~i.~Sekiguchi, M.~Shibata and K.~Taniguchi,
  %``Remnant massive neutron stars of binary neutron star mergers: Evolution process and gravitational waveform,''
  Phys.\ Rev.\ D {\bf 88}, 044026 (2013)
  doi:10.1103/PhysRevD.88.044026
  [arXiv:1307.5888 [astro-ph.HE]].
  
  
  
  %\cite{Sekiguchi:2011zd}
\bibitem{Sekiguchi:2011zd} 
  Y.~Sekiguchi, K.~Kiuchi, K.~Kyutoku and M.~Shibata,
  %``Gravitational waves and neutrino emission from the merger of binary neutron stars,''
  Phys.\ Rev.\ Lett.\  {\bf 107}, 051102 (2011)
%  doi:10.1103/PhysRevLett.107.051102
 % [arXiv:1105.2125 [gr-qc]].
 
 
 
 
 %\cite{Lippuner:2017bfm}
\bibitem{Lippuner:2017bfm} 
  J.~Lippuner, R.~Fernández, L.~F.~Roberts, F.~Foucart, D.~Kasen, B.~D.~Metzger and C.~D.~Ott,
  %``Signatures of hypermassive neutron star lifetimes on r-process nucleosynthesis in the disk ejecta from neutron star mergers,''
  Mon.\ Not.\ Roy.\ Astron.\ Soc.\  {\bf 472}, 904 (2017)
%  doi:10.1093/mnras/stx1987
%  [arXiv:1703.06216 [astro-ph.HE]].
  
%\cite{Totani:1997vj}
\bibitem{Totani:1997vj} 
  T.~Totani, K.~Sato, H.~E.~Dalhed and J.~R.~Wilson,
  %``Future detection of supernova neutrino burst and explosion mechanism,''
  Astrophys.\ J.\  {\bf 496}, 216 (1998)    
  
 %\cite{Nakazato:2013maa}
\bibitem{Nakazato:2013maa} 
  K.~Nakazato,
  %``Imprint of Explosion Mechanism on Supernova Relic Neutrinos,''
  Phys.\ Rev.\ D {\bf 88}, no. 8, 083012 (2013)   
  

  
  %\cite{Hogg:1999ad}
\bibitem{Hogg:1999ad} 
  D.~W.~Hogg,
  %``Distance measures in cosmology,''
  astro-ph/9905116.
  
 % 
\bibitem{Kistler2013}
	Kistler et al., 
	%``The impact of metallicity on the rate of type Ia supernovae,''
	The Astrophysical Journal, 770:88, (2013)
% 	doi:10.1088/0004-637X/770/2/88

%\cite{Kistler:2013jza}
\bibitem{Kistler:2013jza} 
  M.~D.~Kistler, H.~Yuksel and A.~M.~Hopkins,
  %``The Cosmic Star Formation Rate from the Faintest Galaxies in the Unobservable Universe,''
  arXiv:1305.1630 [astro-ph.CO].
 

 %\cite{Woosley:2011aw}
\bibitem{Woosley:2011aw} 
  S.~E.~Woosley and A.~heger,
  %``Long Gamma-Ray Transients from Collapsars,''
  Astrophys.\ J.\  {\bf 752}, 32 (2012)
 % doi:10.1088/0004-637X/752/1/32
%  [arXiv:1110.3842 [astro-ph.HE]]. 
  

%\cite{Nakazato:2012qf}
\bibitem{Nakazato:2012qf} 
  K.~Nakazato, K.~Sumiyoshi, H.~Suzuki, T.~Totani, H.~Umeda and S.~Yamada,
  %``Supernova Neutrino Light Curves and Spectra for Various Progenitor Stars: From Core Collapse to Proto-neutron Star Cooling,''
  Astrophys.\ J.\ Suppl.\  {\bf 205}, 2 (2013)
%  doi:10.1088/0067-0049/205/1/2
%  [arXiv:1210.6841 [astro-ph.HE]].

%\cite{deMink:2015yea}
\bibitem{deMink:2015yea} 
  S.~E.~de Mink and K.~Belczynski,
  %``Merger rates of double neutron stars and stellar origin black holes: The Impact of Initial Conditions on Binary Evolution Predictions,''
  Astrophys.\ J.\  {\bf 814}, no. 1, 58 (2015)
%  doi:10.1088/0004-637X/814/1/58
%  [arXiv:1506.03573 [astro-ph.HE]].
  %%CITATION = doi:10.1088/0004-637X/814/1/58;%%
  %37 citations counted in INSPIRE as of 07 Dec 2017

%\cite{Abbott:2016ymx}
\bibitem{Abbott:2016ymx} 
  B.~P.~Abbott {\it et al.} [LIGO Scientific and Virgo Collaborations],
  %``Upper Limits on the Rates of Binary Neutron Star and Neutron Star–black Hole Mergers From Advanced Ligo’s First Observing run,''
  Astrophys.\ J.\  {\bf 832}, no. 2, L21 (2016)
  doi:10.3847/2041-8205/832/2/L21
  [arXiv:1607.07456 [astro-ph.HE]].
  %%CITATION = doi:10.3847/2041-8205/832/2/L21;%%
  %72 citations counted in INSPIRE as of 07 Dec 2017  

%%%%%%%%%%%%%%%%%%%%%%%%%%%%%%%%%%%%%%%%%%%%%%%%%%%%%%%%%%%%%%%%%%%%%%%%%%%%%%%%


\end{thebibliography}
\end{document}